%% file: main.tex
\documentclass[journal=jpcbfk,manuscript=article,layout=twocolumn]{achemso}

\include{math_header}

\newcommand{\added}[1]{{#1}}

\author{Chatipat Lorpaiboon}
\affiliation[UCChem]{Department of Chemistry, University of Chicago, Chicago, IL 60637}
\altaffiliation{Equal Contributions}
\alsoaffiliation[JFI]{James Franck Institute, University of Chicago, Chicago, IL 60637}
\email{chatipat@uchicago.edu}
\author{Erik Henning Thiede}
\affiliation[Flatiron]{Flatiron Institute, New York, NY 60637}
\altaffiliation{Equal Contributions}
\alsoaffiliation[UCCS]{Department of Computer Science, University of Chicago, Chicago, IL 60637}
\email{thiede@uchicago.edu}
\author{Robert J. Webber}
\altaffiliation{Equal Contributions}
\email{rw2515@nyu.edu}
\affiliation[Courant]{Courant Institute of Mathematical Sciences, New York University, New York, NY 10012}
\author{Jonathan Weare}
\affiliation[Courant]{Courant Institute of Mathematical Sciences, New York University, New York, NY 10012}
\email{weare@cims.nyu.edu}
\author{Aaron R. Dinner}
\affiliation[UCChem]{Department of Chemistry, University of Chicago, Chicago, IL 60637}
\alsoaffiliation[JFI]{James Franck Institute, University of Chicago, Chicago, IL 60637}
\email{dinner@uchicago.edu}

\title[Integrated VAC]{Integrated VAC: A robust strategy for identifying eigenfunctions of dynamical operators}

\begin{document}

%%%%%%%%%%%%%%%%%%%%%%%%%%%%%%%%%%%%%%%%%%%%%%%%%%%%%%%%%%%%%%%%%%%%%
%% The "tocentry" environment can be used to create an entry for the
%% graphical table of contents. It is given here as some journals
%% require that it is printed as part of the abstract page. It will
%% be automatically moved as appropriate.
%%%%%%%%%%%%%%%%%%%%%%%%%%%%%%%%%%%%%%%%%%%%%%%%%%%%%%%%%%%%%%%%%%%%%
% \begin{tocentry}

% Some journals require a graphical entry for the Table of Contents.
% This should be laid out ``print ready'' so that the sizing of the
% text is correct.

% Inside the \texttt{tocentry} environment, the font used is Helvetica
% 8\,pt, as required by \emph{Journal of the American Chemical
% Society}.

% The surrounding frame is 9\,cm by 3.5\,cm, which is the maximum
% permitted for  \emph{Journal of the American Chemical Society}
% graphical table of content entries. The box will not resize if the
% content is too big: instead it will overflow the edge of the box.

% This box and the associated title will always be printed on a
% separate page at the end of the document.

% \end{tocentry}
\begin{tocentry}
\centering
\includegraphics[scale = .8]{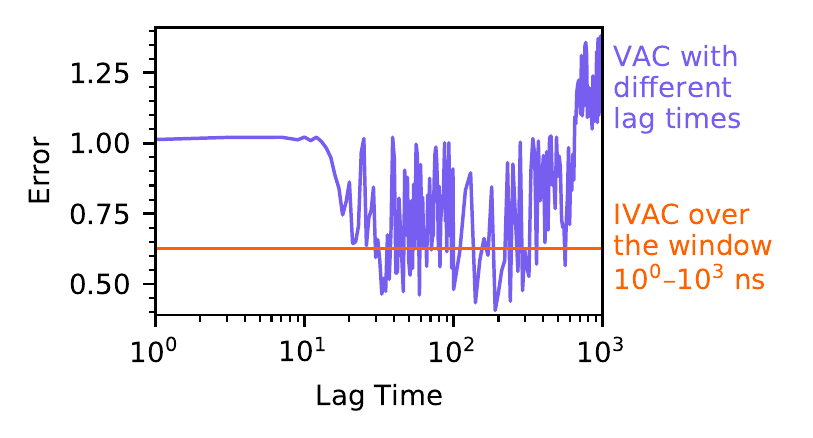}
\end{tocentry}
%%%%%%%%%%%%%%%%%%%%%%%%%%%%%%%%%%%%%%%%%%%%%%%%%%%%%%%%%%%%%%%%%%%%%
%% The abstract environment will automatically gobble the contents
%% if an abstract is not used by the target journal.
%%%%%%%%%%%%%%%%%%%%%%%%%%%%%%%%%%%%%%%%%%%%%%%%%%%%%%%%%%%%%%%%%%%%%
\begin{abstract}
\input{abstract}
\end{abstract}

%%%%%%%%%%%%%%%%%%%%%%%%%%%%%%%%%%%%%%%%%%%%%%%%%%%%%%%%%%%%%%%%%%%%%
%% Start the main part of the manuscript here.
%%%%%%%%%%%%%%%%%%%%%%%%%%%%%%%%%%%%%%%%%%%%%%%%%%%%%%%%%%%%%%%%%%%%%
\section{Introduction}
\input{intro}

\input{background}
\input{iVAC}

\input{alanine_dipeptide}

\input{villin_headpiece}

\section{Conclusion\label{conclusion}}
\input{conclusion}

%%%%%%%%%%%%%%%%%%%%%%%%%%%%%%%%%%%%%%%%%%%%%%%%%%%%%%%%%%%%%%%%%%%%%
%% The "Acknowledgement" section can be given in all manuscript
%% classes.  This should be given within the "acknowledgement"
%% environment, which will make the correct section or running title.
%%%%%%%%%%%%%%%%%%%%%%%%%%%%%%%%%%%%%%%%%%%%%%%%%%%%%%%%%%%%%%%%%%%%%
\begin{acknowledgement}

\input{acknowledgments}
% Please use ``The authors thank \ldots'' rather than ``The
% authors would like to thank \ldots''.

% The author thanks Mats Dahlgren for version one of \textsf{achemso},
% and Donald Arseneau for the code taken from \textsf{cite} to move
% citations after punctuation. Many users have provided feedback on the
% class, which is reflected in all of the different demonstrations
% shown in this document.

\end{acknowledgement}

%%%%%%%%%%%%%%%%%%%%%%%%%%%%%%%%%%%%%%%%%%%%%%%%%%%%%%%%%%%%%%%%%%%%%
%% The same is true for Supporting Information, which should use the
%% suppinfo environment.
%%%%%%%%%%%%%%%%%%%%%%%%%%%%%%%%%%%%%%%%%%%%%%%%%%%%%%%%%%%%%%%%%%%%%
\begin{suppinfo}

Alanine dipeptide simulation details and
error plots for additional individual alanine dipeptide trajectories.
Loss functions for villin nonlinear VAC and IVAC training.
% This will usually read something like: ``Experimental procedures and
% characterization data for all new compounds. The class will
% automatically add a sentence pointing to the information on-line:

\end{suppinfo}

%%%%%%%%%%%%%%%%%%%%%%%%%%%%%%%%%%%%%%%%%%%%%%%%%%%%%%%%%%%%%%%%%%%%%
%% The appropriate \bibliography command should be placed here.
%% Notice that the class file automatically sets \bibliographystyle
%% and also names the section correctly.
%%%%%%%%%%%%%%%%%%%%%%%%%%%%%%%%%%%%%%%%%%%%%%%%%%%%%%%%%%%%%%%%%%%%%
\bibliography{vac}

\end{document}

\beginsupplement

\section{Simulation details for the alanine dipeptide experiments}\label{adp_details}

All simulations were conducted using Gromacs~5.1.4~\cite{abraham2015gromacs,pall2014tackling,pronk2013gromacs,van2005gromacs,lindahl2001gromacs,berendsen1995gromacs}.
The molecule was represented by the CHARMM 27 force field \cite{mackerell2000development}
in a solvent modelled by 513 water molecules using a rigid TIP3P model \cite{berendsen1981interaction}.
Long-range electrostatics were performed using particle-mesh Ewald summation at fourth order with a Fourier spacing of 0.12 nm \cite{essmann1995smooth}.
Each simulation used Langevin dynamics, integrated using the GROMACS leap-frog Langevin integrator with a 1 fs time step and a time constant of 0.1 ps at a temperature of 310 K.
Hydrogen bonds were constrained to be rigid using LINCS \cite{hess1997lincs}, and water rigidity was enforced using SETTLE \cite{miyamoto1992settle}.
In each simulation, the system was initialized at a density of 1 kg / L.
The system was then equilibrated for 50~ps at constant volume, followed by another 50~ps equilibration at constant pressure using the Parrinello-Rahman barostat\cite{parrinello1981polymorphic}.
Finally, the system was again equilibrated at constant volume for 50~ns.
The data set used was obtained from a production run of 50~ns, with structures saved every 500~fs.
To construct our references for the true eigenfunctions, we ran 10 simulations each of length 150~ns, and constructed an MSM on all dihedral angles.
This MSM had 500 Markov states; these were identified by k-means clustering, and we estimated the eigenfunctions and eigenvalues using pyEMMA \cite{scherer2015pyemma}.

\bibliography{vac}

\pagebreak
%\subsection{VAC and IVAC errors for all alanine dipeptide trajectories}

\begin{figure*}
\centering
\includegraphics[width=0.8\textwidth]{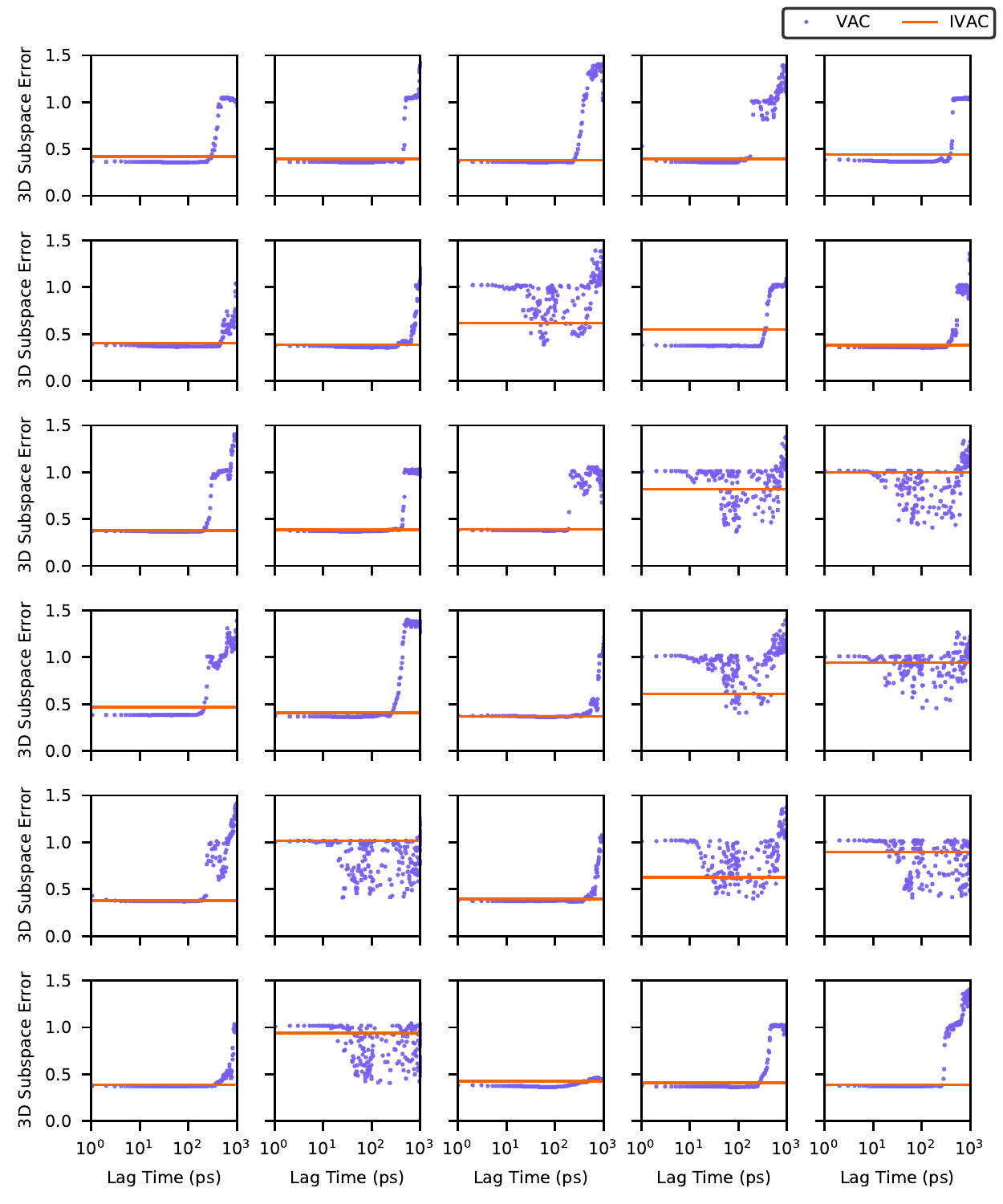}
\caption{VAC (at the horizontal axis lag time) and IVAC (with $\tau_\text{min} = 1~\text{ps}$ and $\tau_\text{max} = 1~\text{ns}$) errors for all 30 of the 10-ns long alanine dipeptide trajectories.}
\end{figure*}

\begin{figure*}
\centering
\includegraphics[width=0.8\textwidth]{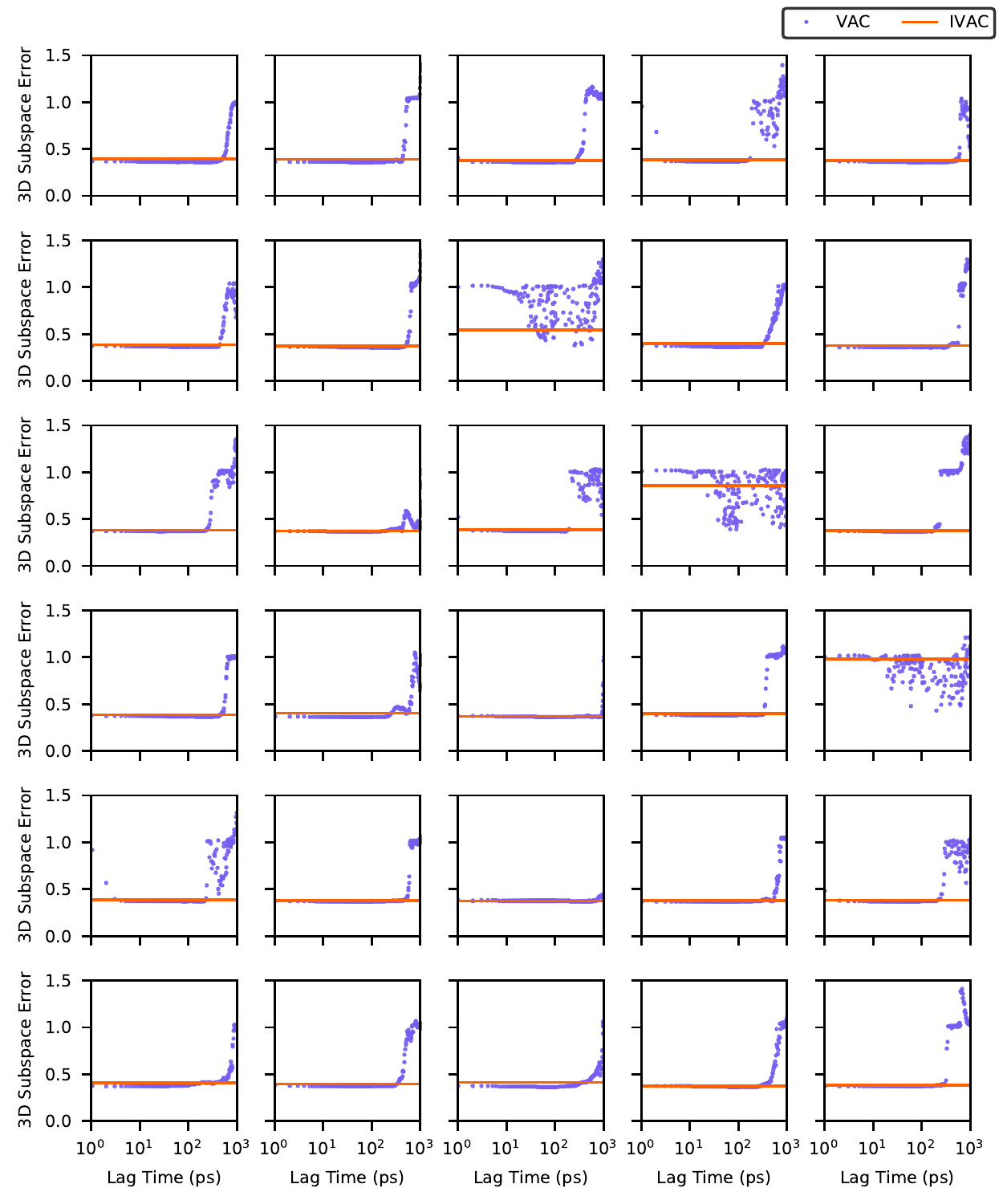}
\caption{VAC (at the horizontal axis lag time) and IVAC (with $\tau_\text{min} = 1~\text{ps}$ and $\tau_\text{max} = 1~\text{ns}$) errors for all 30 of the 20-ns long alanine dipeptide trajectories.}
\end{figure*}

\pagebreak

%\subsection{Nonlinear VAC and IVAC training}

\begin{figure*}
\centering
\includegraphics[width=0.9\textwidth]{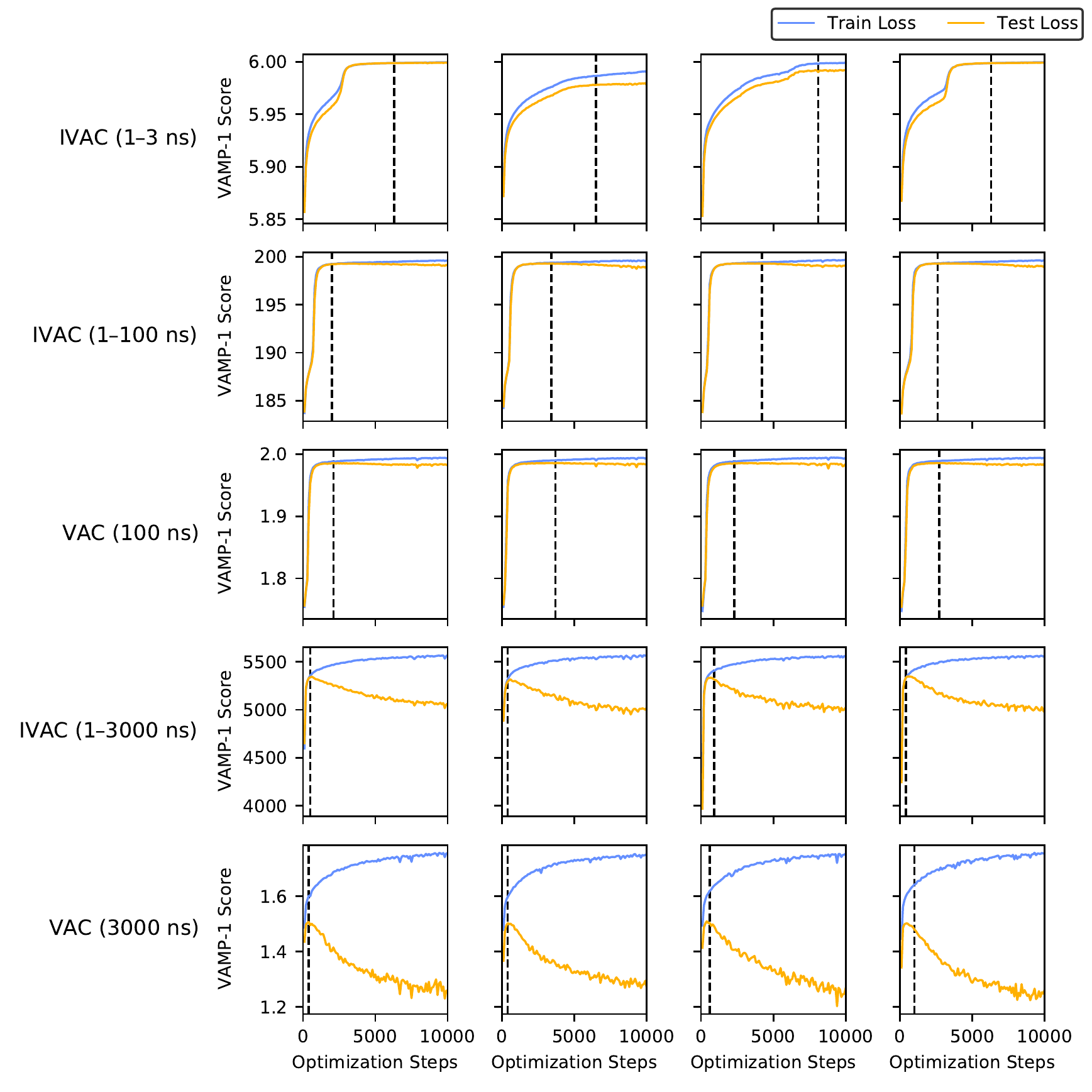}
\caption{
    \added{
        VAMP-1 scores as a function of optimization steps during the training of the neural networks on the villin headpiece data set.
% The central and last rows, VAC (100 ns) and VAC (3000 ns), are equivalent using the VAMP-1 score.
Early-stopping times marked by the black dotted lines.
    }
    }
\end{figure*}

\pagebreak

%% file: math_header.tex
\usepackage{amsmath,amsfonts,amssymb,amsthm,dsfont,xcolor,graphicx,subcaption,listings,dcolumn,mathtools,enumitem}
\usepackage[ruled]{algorithm2e}

\newcommand{\<}{\left \langle}
\renewcommand{\>}{\right \rangle}
\newcommand{\E}{\mathrm{\mathbf{E}}} %true expectation

\newcommand{\T}{\mathcal{T}} %Transition Operator
\newcommand{\its}{\text{ITS}}

\newcommand{\taumin}{{\tau_{\text{min}}}}
\newcommand{\taumax}{{\tau_{\text{max}}}}

\DeclareMathOperator*{\argmax}{arg\,max}
\DeclareMathOperator{\tr}{tr}

%% file: abstract.tex
% One approach to analyzing molecular dynamics data is to search for long-lived patterns in time series of internal coordinates,
% especially slowly decorrelating patterns that indicate large-scale conformational changes.
One approach to analyzing the dynamics of a physical system is to search for long-lived patterns in its motions.
This approach has been particularly successful for molecular dynamics data, where slowly decorrelating patterns can indicate large-scale conformational changes.
Detecting such patterns is the central objective of the variational approach to conformational dynamics (VAC),
as well as the related methods of time-lagged independent component analysis and Markov state modeling.
In VAC, the search for slowly decorrelating patterns is formalized as a variational problem solved by the eigenfunctions of the system's transition operator.
VAC computes solutions to this variational problem
by optimizing a linear or nonlinear model of the eigenfunctions using time series data.
Here, we build on VAC's success by addressing two practical limitations.
First, VAC can give poor eigenfunction estimates when the lag time parameter is chosen poorly.
Second, VAC can overfit when using flexible parameterizations such as artificial neural networks with insufficient regularization.
To address these issues, we propose an extension
that we call integrated VAC (IVAC).
IVAC integrates over multiple lag times before solving the variational problem,
making its results more robust and reproducible than VAC's.

%% file: intro.tex
Many physical systems exhibit motion across fast and slow timescales.
Whereas individual subcomponents may relax rapidly to a quasi-equilibrium, large collective motions occur over timescales that are orders of magnitude longer.
These slow motions are often the most scientifically significant.
For instance, observing the large-scale conformational changes that govern protein function requires microseconds to days, even though individual atomic vibrations have periods of femtoseconds.
However, when exploring new systems, such slow collective processes
may not be fully understood from the outset.
Rather, they must be detected from time series data.

% The rate-determining steps of molecular processes are highly sensitive to dynamical bottlenecks.
% For example,
% while fast processes such as atomic vibrations or water rearrangements relax on recovered from time series datatimescales of picoseconds or faster,
% protein conformations that determine functionality
% typically change on timescales of microseconds or longer.
% It is the slowly-evolving conformational changes that are rate-limiting,
% and so identifying slowly decorrelating functions is of major interest 
% in protein simulations.
% More generally, there is interest in identifying slowly decorrelating functions from molecular dynamics data sets.

One approach for automating this process 
is the ``variational approach to conformational dynamics" (VAC)\cite{noe2013variational, nuske2014variational, perez2013identification}.
In the VAC framework, slow dynamical processes are identified using functions that decorrelate slowly.
These functions are the eigenfunctions of a self-adjoint operator associated with the system's dynamics known as the \textit{transition operator}. 
The transition operator evolves expectations of functions over the system's state forward in time and completely defines the dynamics on a distributional level.
% The transition operator governs the evolution of the expectation of any function of the system's state, thereby completely defining the dynamics on a distributional level.
% The eigenfunctions of the transition operator with the largest eigenvalues govern the system's long-time dynamics.
% More precisely, the subspace spanned by these eigenfunctions contains the functions that decay most slowly to their equilibrium averages. 
% VAC estimates these eigenspaces by constructing a variational ansatz and tuning the ansatz to maximize the time-correlation function.
VAC estimates the transition operator's eigenfunctions by constructing a linear or nonlinear model and using data to optimize parameters in the model.   
VAC encompasses commonly used 
%algorithms 
approaches
such as time-lagged independent component analysis \cite{molgedey1994separation,perez2013identification,schwantes2013improvements,klus2018data} and eigenfunction estimates constructed using Markov state models.\cite{schutte1999direct, swope2004a, swope2004b,klus2018data}
In addition, recent VAC approaches use artificial neural networks to learn approximations to the eigenfunctions.\cite{mardt2017vampnets,chen2019nonlinear}

While VAC has been successful in some applications, 
the approach has limitations.
The accuracy of the estimated eigenfunctions depends strongly on the function space in which the eigenfunctions are approximated, the amount of data available, and a hyperparameter known as the lag time.
In our previous work\cite{vacpaper1} we gave a comprehensive error analysis for the linear VAC algorithm.
This error analysis showed that the choice of lag time can be critical to achieving an accurate VAC scheme.
Choosing a lag time that is too short can cause substantial systematic bias in estimated eigenfunctions, while choosing a lag time that is too long can make VAC exponentially sensitive to sampling error.
% A systematic approach to choosing this hyperparameter is needed.

% Previous work has constructed eigenfunction and implied timescale estimates using information using data from multiple time lags by applying techniques such as
% observable operator models\cite{wu2015projected}, memory kernel models \cite{cao2020advantages}, and delay embedding \cite{thiede2019galerkin}.
% In these approaches, this information was used to account for unobserved degrees of freedom in the system.
% However, we believe the potential ability of information over multiple time lags to improve the statistical quality of eigenfunction estimates has been under-recognized. 
% In contrast, IVAC has a different focus
% since it uses multiple lag tim    s
% to improve the statistical quality of the eigenfunction estimates.

% Previous work has used data from multiple lag times to account for unobserved degrees of freedom, via various techniques including observable operator models\cite{wu2015projected}, memory kernel models\cite{cao2020advantages}, and delay embedding\cite{thiede2019galerkin}.
% %% complementary approach? existing vs recovering information?
% Our work focuses instead on using multiple lag times to reduce the variation in eigenspace estimates due to stochastic noise.

In this paper, we present an extension of the VAC procedure in which we integrate the correlation functions in VAC over a time window.  We term this approach integrated VAC (IVAC).
Because IVAC is less sensitive to the choice of lag time, it reduces error compared to VAC.
Additionally, when IVAC is applied using an approximation space parameterized by a neural network, the approach leads to stable training
and mitigates the overfitting problems associated with VAC.

We organize the rest of the paper as follows.
In the theory section, we review the role of the transition operator and its eigenfunctions, and we introduce the VAC approach for estimating eigenfunctions.  
We then present the procedure for IVAC.
In the results section, we evaluate the performance of IVAC on two model systems.
We conclude with a summary and a discussion of further ways IVAC can be extended.

%% file: background.tex
\section{Methods}
\subsection{Background\label{theory}}

In this section, we review the VAC theoretical framework \cite{takano1995relaxation,noe2013variational} that shows how the slowly decorrelating functions in a physical system can be identified 
using a linear operator
known as the transition operator.

We assume that the system of interest is a continuous-time Markov process $X_{t} \in \mathbb{R}^n$ with a stationary, ergodic distribution $\mu$
(specifically a Feller process \cite{kallenberg2006foundations}).
We use $\E$ to denote expectations of the process $X_t$ started from $\mu$.
For example,  if $\mu$ is the Boltzmann distribution associated with the Hamiltonian $H$ and temperature $T$, then expectations of the process satisfy
\begin{equation}
    \E \left[ f\left(X_{t}\right)\right] = \frac{\int f(x) e^{- H(x) / k_B T} dx }{\int e^{- H(x) / k_B T} dx}
\end{equation}
for all $t \geq 0$.
However, our results are valid for systems with other, more general, stationary distributions.

\subsubsection{The transition operator}\label{ssec:transition_op}

To begin, 
we consider the space of real-valued functions with finite second moment ($\E\left[ f\left( X_0 \right)^2 \right]< \infty$).
Equipped with the inner product
\begin{equation}
    \< f, g \> = \E \left[ f\left( X_0 \right) g\left( X_0 \right) \right]
    \label{eq:defn_inner_product}
\end{equation}
this forms a Hilbert space, which we denote $L^2_\mu$.
We define the transition operator\cite{kallenberg2006foundations} at a lag time $\tau$ to be the operator
\begin{equation}\label{eq:defn_transition_operator}
    \T_\tau f(x) = \E \left[f \left(X_{\tau}\right) | X_{0} = x\right]
\end{equation}
applied to a function $f \in L^2_\mu$.
Here, we are interpreting the conditional expectation as a function of the initial point $x$.
% By an application of Jensen's inequality, it is seen that the function $\T_\tau f$ is in $L^2_\mu$,
% and moreover
% $\T_\tau$ is a linear map from $L^2_\mu$ into $L^2_\mu$.

The transition operator is also called the Markov or (stochastic) Koopman operator \cite{eisner2015operator, klus2018data}.
We use the term transition operator as it is well-established in the literature on stochastic processes, and the terminology emphasizes the connection with finite-state Markov chains.
For a finite-state Markov chain, $f$ is a column vector and $\T_{\tau}$ is a row-stochastic transition matrix.

The transition operator lets us rewrite correlation functions in terms of inner products in $L^2_\mu$:
\begin{align}
     \E\left[ f\left(  X_0 \right) g\left( X_\tau \right) \right] 
    % =&  \E\left[  \E \left[ f\left(  X_0 \right) g\left( X_\tau \right) | X_0 \right] \right] \\
    %&=  \E\left[  f\left(  X_0 \right) \E \left[  g\left( X_\tau \right) | X_0 \right] \right] \\
    &=  \< f, \T_\tau g\>.
    \label{eq:corr_fxn}
\end{align}
% Considering the Rayleigh quotient
Moreover, we can express the slow motions of a system's dynamics
in terms of the transition operator.
The slow motions are identified by functions $f$ for which the normalized correlation function
\begin{equation}
    \frac{\E \left[ f(X_0) f(X_\tau) \right] }{\E \left[ f(X_0) f(X_0) \right] }
    = \frac{\< f, \T_\tau f \> }{\< f, f\>}
    \label{eq:Rayleigh_w_T_op}
\end{equation}
is large.
% suggests an approach for detecting a system's slow motions.
% The slow motions can be identified by functions $f$ for which the Rayleigh quotient is comparatively large.
We will show in the next subsection that these slowly decorrelating functions lie in the linear span of the top eigenfunctions of the transition operator.
% We expect slow motions to correlate strongly with functions that decorrelate slowly,
% causing the left-hand side of~\eqref{eq:Rayleigh_w_T_op} to be large.
% The right hand side suggests that these functions can be characterized through the properties of $\T_\tau$.
% Indeed, we now show that under certain assumptions, the slowest decorrelation functions are the eigenfunctions of the transition operator. 
% immediately suggests an analogy to quantum chemistry.  In the next section  we show that, just as trial wavefunctions can be optimized to obtain estimates of the eigenvalues and eigenfunctions of a quantum mechanical Hamiltonian, the Rayleigh quotient \eqref{eq:Rayleigh_w_T_op} can be maximized to obtain estimates of the eigenvalues and eigenfunctions of the transition operator.

% Forming the ratio (Rayleigh quotient) 
% \begin{equation}
%     \frac{\E \left[ f(X_0) f(X_\tau) \right] }{\E \left[ f(X_0) f(X_0) \right] }
%     = \frac{\< f, \T_\tau f \> }{\< f, f\>}
%     \label{eq:Rayleigh_w_T_op}
% \end{equation}
% immediately suggests an analogy to quantum chemistry.  In the next section  we show that, just as trial wavefunctions can be optimized to obtain estimates of the eigenvalues and eigenfunctions of a quantum mechanical Hamiltonian, the Rayleigh quotient \eqref{eq:Rayleigh_w_T_op} can be maximized to obtain estimates of the eigenvalues and eigenfunctions of the transition operator.

\subsubsection{Eigenfunctions of the transition operator}\label{ssec:vac}
We can immediately see that $\T_\tau$ has the constant function as an eigenfunction, because
\begin{equation}
    \T_\tau 1 = \E \left[ 1 | X_0=x \right] = 1.
    \label{eq:constant_efxn}
\end{equation}
However, there is no guarantee that any other eigenfunctions exist.
We must therefore impose additional assumptions.

We first assume that $X_t$ obeys detailed balance.
For any functions $f,g \in L^2_\mu$, we have
\begin{equation}
    \E \left[ f\left( X_0 \right) g\left( X_\tau \right) \right] = \E \left[ f\left( X_\tau \right) g\left( X_0 \right) \right],
\end{equation}
or equivalently
\begin{equation}
    \< f, \T_{\tau} g\> = \< \T_\tau f, g \>.
\end{equation}
This detailed balance condition ensures that $\T_\tau$ is a self-adjoint operator on $L^2_\mu$.

Next we assume that
$\T_\tau$ is a compact operator.
In our context, 
assuming compactness
is the same as assuming
that the action 
of $\T_\tau$
can be decomposed as an infinite sum
involving eigenfunctions and eigenvalues:
\begin{equation}
    \T_\tau f(x) = \sum_{i=1}^\infty e^{- \sigma_i \tau} \< \eta_i, f(x) \> \eta_i(x).
    \label{eq:spectral_decomposition}
\end{equation}
Our assumption of compactness is made for the sake of simplicity; in fact a weaker assumption of quasi-compactness is sufficient.
We refer the reader to~\citet{vacpaper1} for a more general treatment.

% The functions $\eta_i$ and values 
% $e^{- \sigma_i \tau}$ are the eigenfunctions and eigenvalues of the transition operator.
% The eigenvalues are indexed so that  (counted with multiplicity and sorted by magnitude) of $\T_\tau$.
At all lag times $\tau > 0$,
the function $\eta_i$ is
an eigenfunction
of the transition operator $T^\tau$ with eigenvalue
\begin{equation}
\lambda_i^\tau = e^{-\sigma_i \tau}.
\end{equation}
The eigenvalues are indexed so that
\begin{equation}
    0 = \sigma_1 < \sigma_2 \leq \sigma_3 \leq \cdots
    \label{eq:eigenvalue_bounds}
\end{equation}
and $\lim_{i \rightarrow \infty} \sigma_i = \infty$.
Because the process is ergodic, it is known that the largest eigenvalue $\lambda_1^\tau = 1$
is a simple eigenvalue
and all other eigenvalues are bounded away from $1$.
The particular dependence of the eigenvalues on $\tau$ occurs because 
the transition operator can be  written as 
\begin{equation}
    \T_\tau = e^{\mathcal{L} \tau}, \, \forall \tau \geq 0
    \label{eq:infinitesimal_generator}
\end{equation}
where  $\mathcal{L}$ is an operator known as the infinitesimal generator \cite{kallenberg2006foundations}.
% Here $-\sigma_i$ and $\eta_i$ are the eigenvalues and eigenfunctions of $\A$,
% \eqref{eq:infinitesimal_generator} induces the structure for the eigenvalues and eigenvectors of $\T_\tau$.
% Under the ergodicity assumption, $\T_\tau$ has a nondegenerate largest eigenvalue $\lambda_1^\tau = 1$, which is associated with the eigenfunction $\eta_1(x)=1$. 
% All other eigenvalues are bounded away from $1$, that is,
% \begin{equation}
%     1 = \lambda_1^\tau > \lambda_2^\tau \geq \lambda_3^\tau \geq \cdots  \geq 0,
%     \label{eq:eigenvalue_list}
% \end{equation}
% and $\lim_{i \rightarrow \infty} \lambda_i^\tau = 0$.
We note that it is also common to consider the \emph{implied timescale} (ITS) associated with eigenfunction $i$, defined as
\begin{equation}
    \its_i =  \sigma_i^{-1}.
\end{equation}

% Lastly, the self-adjointness of $\T_\tau$ implies that the eigenfunctions form an orthogonal basis for $L^2_\mu$.
% Consequently, for an arbitrary function in $L^2_\mu$ we can write
% \begin{equation}
%     f(x) = \sum_{i=1}^{\infty} \eta_i(x) \<\eta_i, f\>.
%     \label{eq:proj_on_basis}
% \end{equation}
We can use the eigenvalues and eigenvectors of the transition operator
% to refine our intuition about the normalized autocorrelation function.
to rewrite the normalized correlation function \eqref{eq:Rayleigh_w_T_op}.
Observing that $\T_0 f(x) = f(x)$ and 
substituting \eqref{eq:spectral_decomposition} into the numerator and denominator of \eqref{eq:Rayleigh_w_T_op} gives 
\begin{equation}
\label{eq:maximize}
    \frac{\E \left[ f(X_0) f(X_\tau) \right] }{\E \left[ f(X_0) f(X_0) \right] }
    = \frac{\sum_{i=1}^{\infty} e^{- \sigma_i \tau} \< \eta_i, f\>^2}{\sum_{i=1}^\infty \<\eta_i, f\>^2}.
\end{equation}

We now consider which functions maximize the normalized correlation function.
Applying~\eqref{eq:eigenvalue_bounds}, we find that the normalized correlation function is maximized when we set $f$ to be the constant 
function $f(x)=\eta_1(x)=1$, because
\begin{align}
    \frac{\sum_{i=1}^{\infty} e^{- \sigma_i \tau} \< \eta_i, f\>^2}{\sum_{i=1}^\infty \<\eta_i, f\>^2} \leq& 
    \frac{\sum_{i=1}^{\infty} e^{- \sigma_1 \tau} \< \eta_i, f\>^2}{\sum_{i=1}^\infty \<\eta_i, f\>^2} \\
    =& e^{- \sigma_1 \tau}
\end{align}
for all functions $f \in L^2_\mu$.
If we constrain the search to functions that are orthogonal to $\eta_1$, i.e., functions where
\begin{equation}
    \<\eta_1, f\> = \E\left[ f(x)\right] = 0
\end{equation}
and assume $\sigma_2 > \sigma_3$,
the normalized correlation function is maximized when $f=\eta_2$.
If we constrain $f$ to be orthogonal to both $\eta_1$ and $\eta_2$,
then the next slowest decorrelating function would be $\eta_3$, and so forth.
% Maximizing~\eqref{eq:Rayleigh_w_T_op} is therefore equivalent to identifying the 
% eigenfunctions of the transition operator with the largest eigenvalues, independently of the lag time used.
% % any lag time $\tau$.
\added{
Maximizing the normalized correlation function at any lag time $\tau$
is therefore equivalent to identifying the eigenfunctions of the transition operator.
}

\added{
Because of the connection to slowly decorrelating functions, the eigenfunctions provide 
a natural coordinate system for dimensionality reduction.
The first few eigenfunctions provide a compact representation of all the slowest motions of the system.
Additionally, clustering data based on the eigenfunction values makes it possible to identify metastable states.
}

\subsubsection{The variational approach to conformational dynamics}

The ``variational approach to conformational dynamics" (VAC)
is a procedure for 
identifying eigenfunctions by maximizing the normalized correlation function.
The first eigenfunction, $\eta_1\left(x\right) = 1$, is known exactly and is set to the constant function.
To identify subsequent eigenfunctions, we parameterize a candidate solution
$f$ using a vector of parameters $\theta$.
We then construct an estimate $\gamma_i$
for the $i$th eigenfunction by tuning the parameters to maximize \eqref{eq:Rayleigh_w_T_op}.
% that is,
We set 
$\gamma_i=f_{\theta'}$, where 
\begin{equation}
    \theta'=\argmax_\theta \frac{\E\left[f_\theta\left(X_0\right) f_\theta\left(X_\tau\right) \right]}
    {\E\left[f_\theta\left(X_0\right) f_\theta\left(X_0\right) \right]}
\label{eq:orthogonality_constraints}
\end{equation}
subject to $\< f_\theta, \gamma_j \> = 0$ for all $j < i$. 
In practice, we use empirical estimates of the correlations constructed from sampled data.  
For instance, if our data set consists of a single equilibrium trajectory 
$x_0, x_\Delta, \ldots x_{T-\Delta}$,
we would then construct the estimate
\begin{align}
    & \hat{\E}\left[f\left(X_0\right) g\left(X_\tau\right) \right] = \nonumber \\
    & \frac{\Delta}{T - \tau} 
    \sum_{s=0}^{\frac{T-\Delta-\tau}{\Delta}}
    \frac{f(x_{s\Delta}) g(x_{s\Delta + \tau})
    + f(x_{s \Delta + \tau}) g(x_{s \Delta})}{2}.
    \label{eq:empirical_corr_fxn}
\end{align}
Here and in the rest of the paper, we \added{use the $\,\hat{}\,$ symbol to indicate quantities constructed using sampled data.}

Once we have obtained an estimated eigenfunction $\hat{\gamma}_i$
using data, we can estimate the associated eigenvalue and implied timescale using
\begin{align}
    \hat{\lambda}_i^\tau &= \frac{\hat{\E} \left[ \hat{\gamma}_i(X_0) \hat{\gamma}_i (X_\tau) \right] }{\hat{\E} \left[ \hat{\gamma}_i(X_0) \hat{\gamma}_i(X_0) \right] } \label{eq:VAC_eval_eqn}
    \\
    \hat{\sigma}_i & = -\frac{1}{\tau} \log \hat{\lambda}_i^\tau.
\end{align}
If the sampling is perfect, the variational principle ensures that VAC eigenvalues and VAC implied timescales
are bounded from above by the
true eigenvalues $e^{-\sigma_i \tau}$ and implied timescales $\sigma_i^{-1}$,
and the upper bound is achieved when the VAC eigenfunction is the true eigenfunction $\eta_i$.
However, since the empirical estimate \eqref{eq:VAC_eval_eqn} is used in practice, it is possible to obtain  estimates that exceed the variational upper bound.

The earliest VAC approaches estimated the eigenfunctions of the transition operator by using linear combinations of basis functions $\left\{\phi_i \right\}$, a procedure 
now known as linear VAC.
In linear VAC, the optimization parameters are the unknown linear coefficients $v$, which solve the generalized eigenvalue problem
\begin{equation}
    \hat{C}(\tau) {v}_i = \hat{\lambda}_i^\tau \hat{C}(0) {v}_i,
    \label{eq:linear_vac}
\end{equation}
where
\begin{equation}
    \hat{C}_{jk}(t) = \hat{\E} \left[ \phi_j (X_0) \phi_k (X_t) \right].
\end{equation}
In approaches known as time-lagged independent component analysis \cite{molgedey1994separation} and relaxation mode analysis \cite{takano1995relaxation, hirao1997molecular}, the basis functions $\left\{ \phi_i \right\}$ were chosen to be the system's coordinate axes.
This choice of approximation space is still commonly used to construct collective variable spaces either for analyzing dynamics or for streamlining further sampling.
Markov state models (MSMs) 
provide an alternative approach for estimating eigenfunctions using linear combinations of basis functions
\cite{schutte1999direct, swope2004describing, swope2004b,pande2010everything}.
MSMs can serve as general dynamical models for the estimation of metastable structures and chemical rates \cite{vanden2014transition,noe2014introduction, keller2019markov,pande2010everything}.
% While MSM estimates of 
% allowing estimates to be constructed for quantities such as mean-first-passage times and committors \cite{noe2014introduction, keller2019markov}.
% they can also be used to estimate eigenfunctions and eigenvalues of $\T_\tau$,
When MSMs are applied to estimate eigenfunctions and eigenvalues, the approach
is equivalent to performing linear VAC using a basis of indicator functions on disjoint sets \cite{klus2018data}.
% and estimates of chemical rates are generalized by a closely related formalism \cite{thiede2019galerkin}.

% Subsequent work unified these approaches 
% in the procedure known as VAC \cite{noe2013variational}
% and introduced more general linear basis functions.\cite{vitalini2015basis, boninsegna2015investigating, schwantes2014perspective}.
\citet{noe2013variational} unified the linear VAC approaches
and exploited a general variational principle for identifying eigenvalues and eigenfunctions of the transition operator.
Subsequent work further developed the methodology and introduced more general linear basis functions \cite{nuske2014variational, vitalini2015basis, boninsegna2015investigating, schwantes2014perspective}.
Moreover, it was observed that the general variational principle allows one to model the eigenfunctions using nonlinear approximation spaces
such as the output of a neural network \cite{mardt2017vampnets, chen2019nonlinear}.
\added{
This can lead to  very flexible and powerful approximation spaces.
However, in our experience, the greater flexibility can also lead to overfitting problems 
that need to be addressed through regularization.
}
%that we explore in our results section.

\added{
In a common nonlinear VAC approach, 
a neural network outputs a set of functions $\phi_1, \phi_2, \ldots, \phi_S$
}
that serve as a basis set for linear VAC calculations.
The network parameters are then optimized to maximize the VAMP score \cite{wu2017variational}, which under our assumption of detailed balance can be calculated using
\begin{equation}
    \text{VAMP-}k = \sum_{i=1}^S |\hat{\lambda} _i^\tau|^k.
\end{equation}
The hyperparameter $k$ is typically set to 1 or 2.
% Under detailed balance, the VAMP-1 score coincides with the generalized matrix Rayleigh quotient.\cite{mcgibbon2015variational}
In this paper, we use the VAMP-1 score, 
since we find that it leads to more robust training.
We note that the score function we use is also called the generalized matrix Rayleigh quotient\cite{mcgibbon2015variational}.
% Nonlinear VAC leads to very flexible and powerful approximation spaces.
% However, this greater flexibility can also allow overfitting and unreliable training to occur, as indicated in our numerical experiments below.

\subsubsection{Challenges in VAC calculations}

\begin{figure}
	\centering
	\includegraphics[scale = .45, clip]{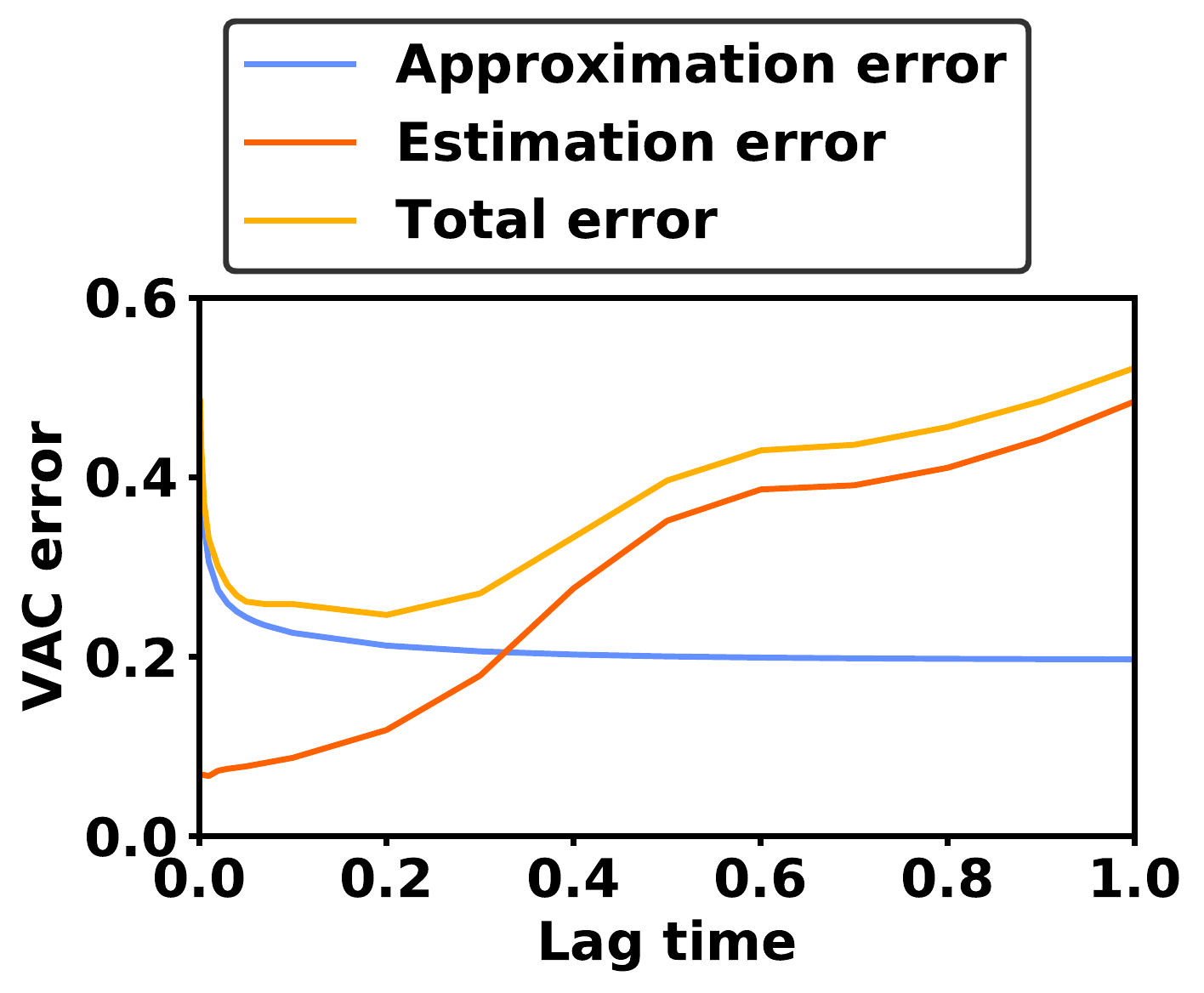}
	\caption{\label{fig:err_decomp} Schematic illustrating the sources of VAC error at different lag times.
	Even without sampling, VAC solutions have approximation error.
	Random variation due to sampling contributes additional estimation error.}
\end{figure}

A major challenge in VAC calculations is selecting the lag time $\tau$.
Since the early days of VAC, it was noted that
lag times that are too short or too long can lead to inaccurate eigenfunction estimates \cite{naritomi2011slow,husic2017note}.
Our recent work \cite{vacpaper1} revealed that the sensitivity to lag time is caused by a combination of approximation error at short lag times
and estimation error at long lag times.
In this section, we describe the impact of approximation error
and estimation error
and provide a schematic (Figure \ref{fig:err_decomp}) that illustrates the tradeoff between approximation error and estimation
error at different lag times.

\emph{Approximation error} is the systematic error of VAC that exists even when VAC is performed with an infinite data set.
We expect approximation error to dominate the calculation when the basis set is of poor quality and our approximation space cannot faithfully represent the eigenfunctions of the transition operator.
The approximation error is greatest at short lag times,
and it decreases and eventually stabilizes as the lag time is increased.
Therefore, VAC users can typically reduce approximation error
by avoiding the very shortest lag times.

% Previous work has attempted to reduce approximation error with a better representation of the system. This includes using data from multiple lag times to account for unobserved degrees of freedom, via various techniques including observable operator models\cite{wu2015projected}, memory kernel models\cite{cao2020advantages}, and delay embedding\cite{thiede2019galerkin}.

\emph{Estimation error} is the random error of VAC that comes from statistical sampling.
As shown in our previous work \cite{vacpaper1}, with increasing lag time the
results of VAC
become exponentially sensitive to small variations in the data set, leading to high estimation error.
At large enough lag times, all the eigenfunction estimates $\hat{\gamma}_2^\tau, \hat{\gamma}_3^\tau, \ldots$ are essentially random noise.

In~\citet{vacpaper1}, we proposed measuring VAC's sensitivity to estimation error
using the condition number $\kappa^\tau$.
The condition number measures the largest possible
changes that can occur in 
the subspace of VAC eigenfunctions 
$\left\{\gamma_j^\tau, \gamma_{j+1}^\tau, \ldots, \gamma_k^\tau\right\}$
when there are small errors in the entries of $C(0)$
and $C(\tau)$.
The condition number is calculated using the expression
\begin{equation}
    \kappa^\tau = \frac{1}{\min\left\{\hat{\lambda}_{j-1}^\tau - \hat{\lambda}_j^\tau,
    \hat{\lambda}_k^\tau - \hat{\lambda}_{k+1}^\tau\right\}}.
    \label{eq:condition_number}
\end{equation}
\added{
For a given problem and a given lag time,
we can use the condition number to determine
which subspaces of VAC eigenfunctions
are highly sensitive to estimation error
and which subspaces are comparatively less sensitive to estimation error.
}
% At long lag times, eigenvalue gaps $\hat{\lambda}_{j-1}^\tau - \hat{\lambda}_j^\tau$ and
% $\hat{\lambda}_k^\tau - \hat{\lambda}_{k+1}^\tau$ approach zero quickly,
% and the condition number quantifies
% the exponential increase in estimation error that can occur.
% As $\tau \rightarrow 0$, the condition number also increases
% yet there is a cancellation of errors in $C\left(0\right)$ and $C\left(\tau\right)$,
% so the estimation error stabilizes.

Although we rigorously derived the condition number only in the case of linear VAC,
we find that the condition number is also helpful for measuring estimation error in nonlinear VAC. 
If $\kappa^\tau \gtrsim 5$ at all lag times $\tau$,
then identifying eigenfunctions is very difficult
and requires a large data set.
We recommend that authors report the condition number
along with their VAC results,
helping readers to assess whether the results are potentially sensitive to estimation error.

%% file: iVAC.tex
\subsection{Integrated VAC}

To address the difficulty inherent in choosing a good lag time,
we propose an extension of VAC called ``integrated VAC" (IVAC)
where we integrate over a range of different lag times
before solving a variational problem.
We find that the new approach
is more robust to lag time selection 
and it often gives better results overall.

Just as VAC maximizes the correlation function in~\eqref{eq:Rayleigh_w_T_op},
IVAC solves a variational problem
by identifying a subspace of functions $f$ that maximize
the integrated correlation function
\begin{equation}
    \int_{\tau_{min}}^{\tau_{max}} \frac{\E \left[ f(X_0) f(X_s) \right] }{\E \left[ f(X_0) f(X_0) \right] } \mathop{ds}.
    \label{eq:int_Rayleigh_w_T_op}
\end{equation}
As in VAC, the functions solving the variational problem
are the eigenfunctions of the transition operator.
When the eigenfunction $\eta_i$ is substituted into the integrated correlation function \eqref{eq:int_Rayleigh_w_T_op},
the resulting expression is related to the implied timescales by
\begin{align}
    & \int_{\tau_{min}}^{\tau_{max}} \frac{\E \left[ \eta_i(X_0) \eta_i(X_s) \right] }{\E \left[ \eta_i(X_0) \eta_i(X_0) \right] } \mathop{ds} \nonumber \\
    &\hspace{6em} = \frac{e^{-\sigma_i \taumin} - e^{- \sigma_i \taumax}}{\sigma_i}.
    \label{eq:ivac_eigenvalue}
\end{align}
Therefore, like VAC, IVAC is a variational approach for identifying both eigenfunctions and implied timescales.

IVAC is a natural extension of VAC;
in the limit as $\taumax$ approaches $\taumin$,
IVAC gives the same eigenfunction and implied timescale estimates as regular VAC.
However, when
$\taumax$ and $\taumin$ are separated from each other,
the results of IVAC and VAC start to diverge.
We find that IVAC  with minimal tuning performs comparably to VAC with optimal tuning.
IVAC has the desirable feature
that it is not very sensitive to the values of $\taumin$ and $\taumax$.

Previous approaches for estimating eigenfunctions using multiple time lags have attempted to reduce approximation error by accounting for unobserved 
degrees of freedom\cite{wu2015projected,suarez2016accurate,cao2020advantages,thiede2019galerkin}.
In contrast, IVAC uses multiple time lags to reduce estimation error and improve robustness to parameter choice.

\subsubsection{Linear IVAC}

Linear IVAC
uses linear combinations of basis functions to maximize the integrated autocorrelation function \eqref{eq:int_Rayleigh_w_T_op}.
However, as simulation data are sampled at discrete
time points, we cannot directly calculate the integral.
We therefore replace \eqref{eq:int_Rayleigh_w_T_op}
with a discrete sum taken over uniformly spaced lag times.
We seek to maximize
\begin{align}
    \sum_{\tau=\taumin}^\taumax
    \frac{\E \left[ f(X_0) f(X_\tau) \right] }{\E \left[ f(X_0) f(X_0) \right] },
    \label{eq:riemannsum}
\end{align}
where $\tau = \taumin,\  \taumin+\Delta,\ \taumin+2\Delta,\ldots,\ \taumax$ and $\Delta$ is the sampling interval.  The discrete sum \eqref{eq:riemannsum} approximates \eqref{eq:int_Rayleigh_w_T_op} up to a constant multiple,
and its value is maximized when $f$ lies within the span of the top eigenfunctions of the transition operator.
Setting $f$ to be the eigenfunction $\eta_i$, we can sum the resulting finite geometric series:
\begin{align}
& \sum_{\tau=\taumin}^\taumax
    \frac{\E \left[ \eta_i(X_0) \eta_i(X_\tau) \right] }{\E \left[ \eta_i(X_0) \eta_i(X_0) \right] } \nonumber \\
&\hspace{4em}= \frac{e^{-\sigma_i \taumin} - e^{-\sigma_i \left(\taumax + \Delta\right)}}{1 - e^{-\sigma_i \Delta}}. \label{eq:ivac_its_impl}
\end{align}

In linear IVAC, we optimize linear combinations of basis functions $\left\{\phi_i \right\}$ to maximize the functional \eqref{eq:riemannsum}.
The optimization parameters 
are the unknown linear coefficients $v$, which solve the generalized eigenvalue problem
\begin{equation}
\label{eq:ivac_eqn}
    \hat{I}(\taumin, \taumax) {v}_i = \hat{\lambda}_i \hat{C}(0) {v}_i,
\end{equation}
where we have defined
\begin{align}
    \hat{C}_{jk}(t) &= \hat\E \left[ \phi_j (X_0) \phi_k (X_t) \right] \\
    \hat{I}\left(\taumin, \taumax\right) &= \sum_{\tau=\taumin}^\taumax \hat{C}(\tau).
\end{align}
We solve the generalized eigenvalue problem to obtain estimates $\hat{\gamma}_i$ for the transition operator's eigenfunctions.
Then,
we form the sum
\begin{equation}
    \sum_{\tau=\taumin}^\taumax
    \frac{\hat{\E} \left[ \hat{\gamma}_i(X_0) \hat{\gamma}_i(X_\tau) \right] }{\hat{\E} \left[ \hat{\gamma}_i(X_0) \hat{\gamma}_i(X_0) \right] },
\end{equation}
and we estimate implied timescales by
solving \eqref{eq:ivac_its_impl} for $\hat{\sigma}_i$ using a root-finding algorithm.

\subsubsection{Nonlinear IVAC}

Nonlinear IVAC maximizes the integrated correlation function \eqref{eq:int_Rayleigh_w_T_op} by constructing approximations in a nonlinear space of functions, for example, those represented by a neural network. 
Specifically, the nonlinear model provides 
a set of functions $\phi_1, \phi_2, \ldots, \phi_S$
that serve as a basis set for linear IVAC.
\added{
The parameters are trained to maximize the VAMP-$k$ score 
% using our approximation to the eigenvalues of $\int \T_\tau$, 
\begin{equation}
    % \text{IVAC-}k = \sum_{i = 1}^S \lvert\hat{\lambda}_i\rvert^k.
    \sum_{i = 1}^S \lvert\hat{\lambda}_i\rvert^k
\end{equation}
where the eigenvalues $\hat{\lambda}_i$ are defined using equation \eqref{eq:ivac_eqn}.
In a linear approximation space, all values of VAMP-$k$ scores lead to identical eigenfunction estimates.
In a nonlinear approximation space,
it is theoretically possible that minimizing with different values of $k$ scores would lead to different estimates.
However,
in practice we find there is little difference between estimates at the minima.
We present our results using $k=1$ because it leads to the most stable convergence; we found that higher values of $k$ are prone to large gradients and, in turn, unstable training.
When $k=1$, the score function can be computed using
}
% The parameters are trained to maximize the IVAC-$k$ score,
% which we define in analogy to the VAMP-$k$ score to be
% \begin{equation}
%     \text{IVAC-}k = \sum_{i = 1}^S \lvert\hat{\lambda}_i\rvert^k.
% \end{equation}
% In a linear approximation space, all IVAC\nobreakdash-$k$ scores lead to identical eigenfunction estimates.
% In a nonlinear approximation space,
% it is theoretically possible that minimizing different IVAC-$k$ scores leads to different estimates.
% However,
% in practice we find there is little difference between estimates at the minima.
% We present our results in terms of IVAC-1 because it leads to the most stable convergence; we found that higher values of $k$ are prone to large gradients and, in turn, unstable training.
% The IVAC-1 score can be computed using
\begin{align}
\label{eq:ivac_objfn}
    % \sum_{\tau=\taumin}^{\taumax} \hat{\lambda}_i^\tau
    % &= \sum_{\tau = \taumin}^{\taumax} \tr\left( \hat{C}(0)^{-1} \hat{C}(\tau) \right) \nonumber \\
    \tr( \hat{C}(0)^{-1} \hat{I}(\taumin, \taumax) ).
\end{align}

The main practical challenge in an application of nonlinear IVAC
is that the basis functions $\phi_1, \phi_2, \ldots, \phi_S$ change
at every iteration, requiring costly re-evaluation of $\hat{C}(0)$,
$\hat{I}(\taumin, \taumax)$, and the gradient
of \eqref{eq:ivac_objfn} with respect to the parameters.
To reduce this cost, we have developed the batch subsampling approach described in Algorithm \ref{alg:subsampling},
which we apply at the start of each optimization iteration.

\begin{algorithm}
\SetInd{0em}{.75em}
\SetKwInOut{Input}{input}
\SetKwInOut{Output}{output}
\Input{data $x_0, \ldots, x_{T-\Delta}$, $\taumin$, $\taumax$, number of samples $N$}
\For{$n \in 1, 2, \ldots, N$}{
  Sample $\tau_n$ from $\{\taumin, \ldots, \taumax \}$\;
  Sample $s_n$ from $\{0, \ldots, T - \tau_n - \Delta\}$\;
}
\Output{sampled pairs $(x_{s_n}, x_{s_n + \tau_n})$}
\caption{subsampling routine \label{alg:subsampling}}
\end{algorithm}
In the subsampling approach, we draw a randomly chosen set of data points, which allow us to estimate the matrix entries $\hat{C}_{ij}(0)$ using
\begin{equation} 
    \sum_{n=1}^N \frac{\phi_i(x_{s_n}) \phi_j(x_{s_n}) + \phi_i(x_{s_n + \tau}) \phi_j(x_{s_n + \tau})}{2N}
\end{equation}
and the matrix entries $\hat{I}_{ij}(\taumin, \taumax)$ using
\begin{equation}
    \sum_{n=1}^N \frac{\phi_i(x_{s_n}) \phi_j(x_{s_n + \tau_n}) + \phi_i(x_{s_n + \tau_n}) \phi_j(x_{s_n})}{2N \Delta / \left(\taumax - \taumin + \Delta\right)}.
\end{equation}

After constructing these random matrices, we calculate the score function~\ref{eq:ivac_objfn}.
% , which we rewrite as
% \begin{equation}
%     \tr\left(\hat{C}(0)^{-1} \hat{I}\left(\taumin, \taumax\right)\right). \label{eq:ivac_objfn}
% \end{equation}
We then use automatic differentiation to obtain
the gradient of the score function
with respect to the parameters,
and we perform an optimization step.
By randomly drawing new data points at each optimization step,
we ensure a thorough sampling of the data set
and we are able to train the nonlinear
representation at reduced cost.
Typically, we find that $10^3$--$10^4$ data points per batch is enough for the score function~\eqref{eq:ivac_objfn} to be estimated with low bias.

%% file: alanine_dipeptide.tex
\section{Results and discussion\label{results}}

% In this section, we provide numerical evidence that IVAC performs about as well as VAC but is much more robust to the choice of parameters.  First, we show results for the alanine dipeptide. VAC performs well on this system owing to ...[SAY WHY]..., so there is little room for improvement; nevertheless, we show that IVAC is much less sensitive to hyperparameter choices.  Second, we show results for the villin headpiece.  Because this system ...[DESCRIBE PROPERTIES THAT MAKE IT HARDER], VAC ...[FILL IN].... We show that....

In this section, we provide evidence that IVAC is more robust than VAC and
can give more accurate eigenfunction estimates.
First, we show results from applying IVAC and VAC to the alanine dipeptide. 
VAC can provide accurate eigenfunction estimates
for this test problem
owing to the large spectral gap
and the approximation space that overlaps closely with the eigenfunctions of the transition operator.
However, VAC requires a careful tuning of the lag time.
In contrast, IVAC is much less sensitive to lag time choice.
IVAC gives solutions that are comparable to VAC with the optimal lag time parameter and substantially better than VAC with a poorly chosen lag time.

Second, we show results for the villin headpiece protein. 
Because the data set has a small number of independent samples and the neural network approximation space is flexible and prone to overfitting,
VAC and IVAC suffer from estimation error at long lag times. 
Despite these challenges, we present a robust protocol for choosing parameters in IVAC to limit the estimation error, and we show that IVAC is less sensitive to overfitting for this problem
compared with VAC.

\subsection{Application to the alanine dipeptide \label{sub:adp}}

In this section we compare linear IVAC and VAC
applied to Langevin dynamics simulations of the alanine dipeptide (i.e., $N$-acetyl-alanyl-$N'$-methylamide) in aqueous solvent;
further simulation details are given in the supporting information.

The alanine dipeptide is a well-studied model for conformational changes in proteins.
Like many protein systems, the alanine dipeptide has dynamics that are dominated by transitions between metastable states.
The top eigenfunctions are 
useful for locating barriers between states,
as these eigenfunctions
change sharply when passing from one well to another.
We focus on estimating $\eta_2$ and $\eta_3$, as
large changes in these eigenfunctions correspond to transitions over the alanine dipeptide's two largest barriers.
\added{
We refer to the span of $\eta_1$, $\eta_2$, and $\eta_3$ as the 3D subspace.
}

In our experiments, we consider trajectories of length $10\ \text{ns}$ and $20\ \text{ns}$.
The trajectories are long enough to observe approximately 15 or 30 transitions respectively along the dipeptide's slowest degree of freedom.
% This is a scientifically relevant data range.
Folding simulations of proteins, such as the villin headpiece considered below, often have a similar number of transitions between the folded and unfolded states.

There are several features that make it possible for VAC to perform well on this example.
% First, the linear approximation space is small (just $9$ basis functions), and it is known to overlap heavily with the top eigenfunctions of the dynamics.
% We are estimating eigenfunctions by using linear combinations of sines and cosines for all the dihedral angles in the molecular backbone.
\added{
First, the linear approximation space, which consists of all the dihedral angles in the molecular backbone, is small (just $9$ basis functions), and it is known to overlap heavily with the top eigenfunctions of the dynamics.
}
Second, we 
% concentrate on estimating the subspace spanned by eigenfunctions $\eta_1$, $\eta_2$, and $\eta_3$.
are estimating a well-conditioned subspace
with a minimum condition number of just
\begin{equation}
\min_{\tau} \kappa^\tau =
\min_{\tau} \left(\hat{\lambda}_3^\tau - \hat{\lambda}_4^\tau\right)^{-1} = 1.4,
\end{equation}
and therefore we do not expect a heavy amplification of sampling error that degrades eigenfunction estimates.
% Third, we use moderately long trajectories of $10-20\text{ns}$, which also helps to cut down on
% estimation error.

To evaluate the error in our eigenfunction estimates,
we compare to ``ground truth" eigenfunctions
computed using a Markov state model
built with a very long time series (1.5\ $\mu\text{s}$) and a fine discretization of the dihedral angles.
We measure error using the projection distance \cite{edelman1998geometry},
which evaluates the overlap between one subspace and the orthogonal complement of another subspace.
For subspaces $\mathcal{U}$ and $\mathcal{V}$ with orthonormal basis functions $\left\{u_i\right\}$ and $\left\{v_i\right\}$, the projection distance is given by
\begin{equation}
    d \left(\mathcal{U}, \mathcal{V}\right) =
    \sqrt{\sum_{i,j} \left(\delta_{ij} - \left<u_i, v_j\right>^2\right)}.
\end{equation}
\added{
This measure, which combines the error in the different eigenfunctions into a single number, is useful because VAC is typically used to identify subspaces of eigenfunctions rather than individual eigenfunctions. The maximum possible error when estimating $k$ eigenfunctions is $\sqrt{k}$.
}

% Moreover, since VAC is commonly used to find a collective variable space, it is appropriate to quantify the error in VAC using a distance between subspaces.
% Moreover, considering the error in estimating the space spanned by the eigenfunctions reflects how VAC is commonly used:
% as a method for finding a collective variable \emph{space}.

\begin{figure}
	\centering
	\includegraphics{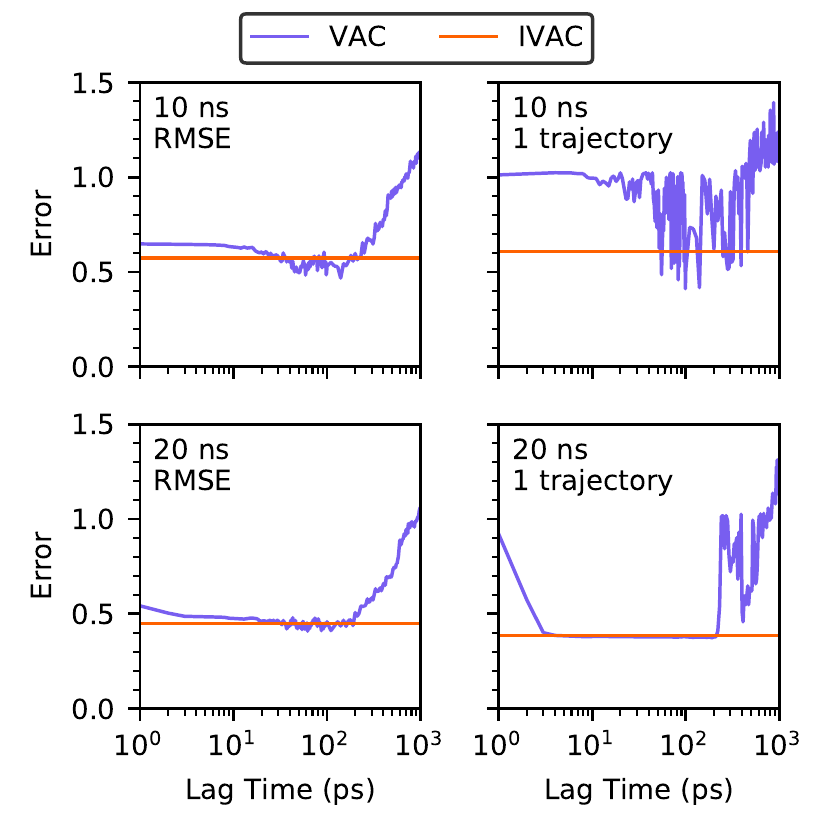}
    \caption{Linear IVAC and VAC errors for alanine dipeptide trajectories.
    IVAC was applied with $\taumin = 1~\text{ps}$ and $\taumax = 1~\text{ns}$.
    VAC was applied with variable lag time $\tau$ (horizontal axis).
    \added{
    Errors are computed using the projection distance to the MSM reference for the span of $\eta_2$ and $\eta_3$.
    (left) Root mean square errors (RMSE) over 30 independent trajectories.
    (right) Errors for a single trajectory.
    }
    }
	\label{fig:adp_data}
\end{figure}

Our main result from the
alanine dipeptide application is that IVAC is more robust to the selection
of lag time parameters than VAC.
In Figure \ref{fig:adp_data}, 
we report the accuracy of IVAC and VAC for different lag times
and trajectory lengths.
In the left column, we show the root mean square errors (RMSE) for IVAC (orange) and VAC (purple), aggregated over thirty independent trajectories.
From the aggregated results, IVAC performs nearly as well
as VAC with the best possible $\tau$
and consistently gives results much better than VAC with a poorly chosen $\tau$.
The RMSE of IVAC is just $0.58$ with $10 \ \text{ns}$ trajectories
and $0.45$ with $20 \ \text{ns}$ trajectories.
These low error levels are not far from the minimum error of $0.37$ that is possible using our linear approximation space.

\begin{figure*}
    	\centering
	\includegraphics{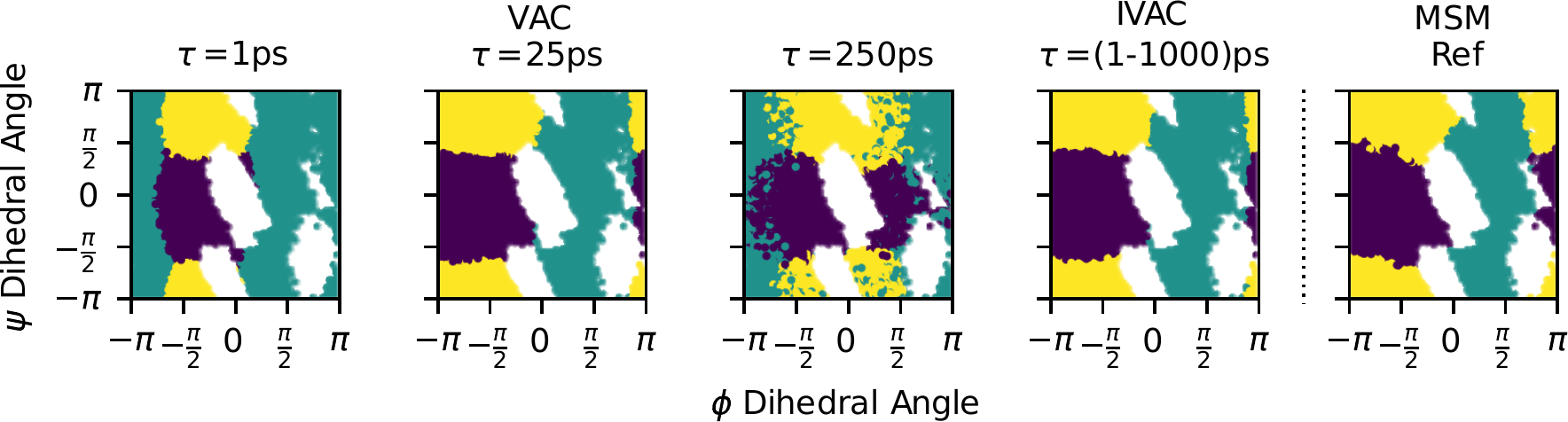}
	\caption{
        \added{
        Clusters on the eigenfunctions estimated using VAC and IVAC compared with clusters on an accurate MSM.  (left of the dashed line) VAC and IVAC results for the 20~ns trajectory from Figure~\ref{fig:adp_data}.  (right of the dashed line) Clustering on $\eta_2$ and $\eta_3$ evaluated using an accurate MSM reference.
        }
	}
   	\label{fig:adp_dpct}
\end{figure*}

\begin{figure*}
	\centering
	\includegraphics{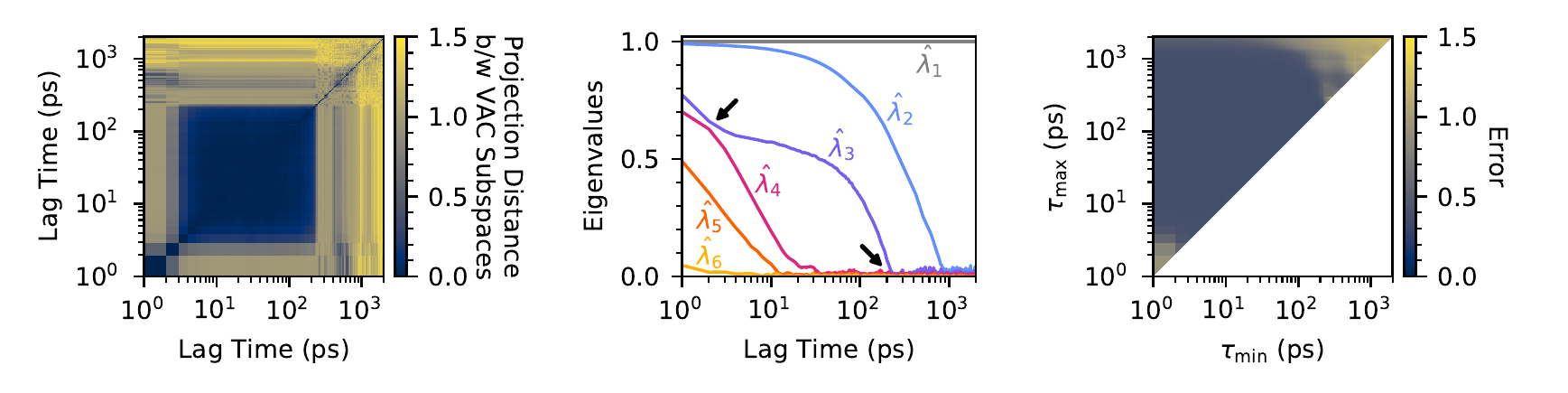}
	\caption{
        \added{
	Lag time dependence of VAC and IVAC results.
	%Consistency of VAC and IVAC results.
	All results shown are for the single 20~ns alanine dipeptide trajectory in Figure~\ref{fig:adp_data}.
	% something about 3D subspace here
	(left) Projection distance between
	VAC results at the horizontal axis lag time
	and VAC results at the vertical axis lag time.
	(center) First six estimated eigenvalues of the transition operator.
	(right) Error in IVAC results at different values of $\taumin$ and $\taumax$,
	evaluated using the projection distance.
    }
    % 	This is the projection distance to the MSM reference.
	%All results shown are for a 20-ns trajectory.
	}
   	\label{fig:adp_select}
\end{figure*}

In the right column of Figure \ref{fig:adp_data}, we show results for a $10 \ \text{ns}$ trajectory and a $20 \ \text{ns}$ trajectory.
The trajectories were selected to help illustrate differences in the error profiles for VAC and IVAC; similar plots for all other trajectories can be found in the supporting information.
We observe two key differences.
First, VAC error can exhibit high-frequency stochastic variability as a function of lag time,
a source of variability that does not affect integrated VAC results.
Second, VAC can have high error levels 
%(higher even than $1.0$)
at very short and long lag times.
The projection \added{distance} against our reference often reaches 1.0, which might indicate that a true eigenfunction is completely orthogonal to our estimated subspace.
The error of IVAC is unlikely to reach such high extremes.

We note that the parameter values $\tau_{\min} = 1\ \text{ps}$ and $\tau_{\max} = 1\ \text{ns}$ used in IVAC are not hard to tune.
The range $1 \ \text{ps} - 1 \ \text{ns}$ is a broad window 
of lag times over which VAC eigenvalues
$\hat{\lambda}_2^\tau$ and $\hat{\lambda}_3^\tau$
decrease from values near one
to values near zero.
In contrast, it is much harder to tune the VAC lag time $\tau$.
VAC results are very sensitive to high or low lag times
as seen in Figure \ref{fig:adp_data}.

% To visualize our results, we clustered the trajectory data using the IVAC and VAC estimates.
\added{
When eigenfunction estimates are accurate, we expect that the 
eigenfunction coordinates will help identify the system's metastable states.  
In Figure~\ref{fig:adp_dpct}, we 
compare the results of clustering configurations in the $20 \ \text{ns}$ alanine dipeptide trajectory in Figure 2 using the associated IVAC and VAC estimates.
We 
plot the predicted metastable states against the dipeptide's $\phi$ and $\psi$ dihedral angles.
In the figure, we present VAC results taken at a short lag time, an intermediate lag time, and a long lag time.
We also present results
% For comparison, we provide clusters 
% constructed directly on the the eigenfunctions 
for the MSM reference.
% Here the VAC lag times have been chosen to depict the three error regimes seen in Figure~\ref{fig:adp_data}. 
% Compared to the reference, we see that VAC performs equally well to IVAC at well-chosen lag times but otherwise performs more poorly.
Comparing against the reference, we find that IVAC identifies clusters as accurately as VAC at a well-chosen lag time, and IVAC performs far better than VAC at a poorly-chosen lag time.
}

Next, we 
present additional analyses 
applied to a single $20 \ \text{ns}$ alanine dipeptide trajectory,
that provide insight into why IVAC
is more robust to lag time selection than VAC.
% We start by reporting internal consistency results for VAC
% in Figure \ref{fig:adp_select}, left.
% To generate the figure,
\added{
To start, we examine the discrepancy in VAC results at different lag times.
In Figure~\ref{fig:adp_select}, left,
}
we performed VAC with a range of different lag times,
and we measured the projection \added{distance} between the VAC results
obtained at one lag time $\tau_1$ (horizontal axis) and the VAC results obtained at a different lag time $\tau_2$ (vertical axis).
% The large dark blue square indicates that VAC's results are internally consistent for lag times $\tau$ chosen within the range of $3\ \text{ps} - 200\ \text{ps}$,
% but they are inconsistent when the lag time is taken outside this range (yellow end of the distance scale).
The square with low projection distance between $3 \ \text{ps}$ and $200 \ \text{ps}$ indicates that VAC results with lag times chosen within this range are similar to one another, but not to those with lag times taken from outside this range.

\added{The discrepancy between VAC results at both} low and high lag times can be explained by a plot of VAC eigenvalues (Figure \ref{fig:adp_select}, center).
At $3\ \text{ps}$, there is an eigenvalue crossing between the eigenvalues $\hat{\lambda}_3^\tau$ and $\hat{\lambda}_4^\tau$ (shown in purple and magenta).
The eigenvalue crossing causes VAC to misidentify the third VAC eigenfunction (which is inside the 3D subspace) and the fourth VAC eigenfunction (which is outside the 3D subspace).
At $200\ \text{ps}$, there is a different problem
related to insufficient sampling.
The third eigenvalue descends into noise,
causing VAC to fit the
first two eigenfunctions
%2D subspace 
at the expense of the 3D subspace.

With integrated VAC, the problem of finding a single good lag time is replaced with the problem of finding two endpoints for a range of lag times.
This proves to be an easier task as IVAC is more tolerant of lag times outside the region where VAC gives good results.
In Figure \ref{fig:adp_select}, right,
we show the error of IVAC as a function of $\tau_{\min}$ and $\tau_{\max}$ (horizontal and vertical axes, respectively).
This figure, which shows the error of IVAC estimates computed from comparison with the reference, is different from the figure on the left which shows only the discrepancy between VAC results at different lag times.
Figure \ref{fig:adp_select}, right, also shows the error of VAC,
which appears along the diagonal of the plot
corresponding to the case $\tau_{\min} = \tau_{\max}$.

Figure \ref{fig:adp_select}, right, reveals that the range of lag time parameters for which IVAC exhibits low error levels
is much broader than the range of lag times for which VAC exhibits low error levels.
This supports our basic argument that choosing good parameters in IVAC is easier than choosing good parameters in VAC.
To achieve low errors, we do not need to identify the optimal VAC lag times but only integrate over a window that contains the optimal VAC lag times while ensuring that $\taumax$ is not excessively high.
% there are enough good lag times included within the averaging window.
% If $\taumin$ is set to $1\ \text{ps}$, which is far too low a lag time for VAC (compare Figure \ref{fig:adp_data}), 
% values of $\taumax$ from $4\ \text{ps}$ to $1\ \text{ns}$ will give a reasonable answer.
% Likewise, if $\taumax$ is set to $1\ \text{ns}$, even decreasing $\taumin$ to $0\ \text{ns}$ will give a reasonable answer.
% The contamination due to bad lag times is mitigated by the presence of good lag times,
% causing IVAC to be robustly effective for this system.

%% file: villin_headpiece.tex
\subsection{Application to the villin headpiece \label{sub:villin}}

Next we apply IVAC to a difficult
spectral estimation problem with limited data.
We seek to estimate the slow dynamics
% corresponding to folding and unfolding transitions
for an engineered 35-residue subdomain of the villin headpiece protein.
Our data consist of a $125\ \mu \text{s}$ molecular dynamics simulation performed by \citet{lindorff2011fast}
Villin is a common model system for protein folding for both experimental and computational studies \cite{mcknight1996thermostable, kubelka2003experimental, duan1998pathways, lindorff2011fast}, 
where
the top eigenfunctions correlate with the folding and unfolding of the protein.

On the surface, the villin data set would seem to be much larger and more useful for spectral estimation compared to the $10-20\ \text{ns}$ trajectories
we examined for the alanine dipeptide.
However, the villin headpiece relaxes to equilibrium orders of magnitude more slowly than the alanine dipeptide.
The data set contains just 34 folding/unfolding events
with a folding time of $2.8\ \mu \text{s}$.
The limited number of observed events is characteristic of 
simulations of larger and more complex biomolecules,
since simulations require massive computational resources
and
conformational changes take place slowly over many molecular dynamics time steps.
\added{
The fact that dynamics of villin are not understood nearly as well as the dynamics of the alanine dipeptide presents an additional challenge.
Compared to the alanine dipeptide, villin has a more complex free energy surface and a larger number of degrees of freedom.
Since the true eigenfunctions of the system are unknown,
it is appropriate to apply spectral estimation using a large and diverse feature set.
However, the large size and diversity of the feature set increases the risk of estimation error.
}

In contrast to the alanine dipeptide results,
where we applied IVAC using linear combinations of basis functions,
here we apply IVAC using a neural network.
The increased flexibility of the neural network approximation reduces approximation error.    
However, the procedure for optimizing the neural network is more complicated than the procedure for applying linear VAC.
Moreover, the %increased 
complexity of the neural network representation 
(around $5 \times 10^4$ parameters)
makes overfitting a concern for this example.

% We use the same neural network as in~\citet{sidky2019trpcage}, with 2 hidden layers of 50 neurons, tanh nonlinearities, and batch normalization between layers.
We use a slight modification of the neural network architecture published in~\citet{sidky2019trpcage}, with 2 hidden layers of 50 neurons,
%(Sidky and coauthors used 100 neurons per layer), 
tanh nonlinearities, and batch normalization between layers.
The network is built on top of a rich set of features, consisting of all the $\text{C}_\alpha$ pairwise distances as well as sines and cosines of all dihedral angles.
At each optimization step, we subsample $10^4$ data points using Algorithm \ref{alg:subsampling}.
We optimize the neural network parameters using AdamW \cite{loshchilov2017decoupled} with a learning rate of $10^{-4}$ and a weight decay coefficient of $10^{-2}$.
Following standard practice, we use the first half of the data set for training
and the second half for validation.
We validate the neural network against the testing data set every 100 optimization steps, and perform early stopping with a patience of 10.
% , although we find that the training is largely insensitive to the patience parameter.
% We perform early stopping at the optimization step that gives the best validation score.
% We set the early stopping patience to 10 epochs, although we find that training is largely insensitive to the patience parameter.

We present our results for villin in two parts.
First we describe our procedure for selecting parameters in nonlinear IVAC.
Next we highlight evidence that nonlinear IVAC shows greater
robustness to overfitting compared to nonlinear VAC.

\subsubsection{Selection of parameters}

\begin{figure*}
\centering
\includegraphics{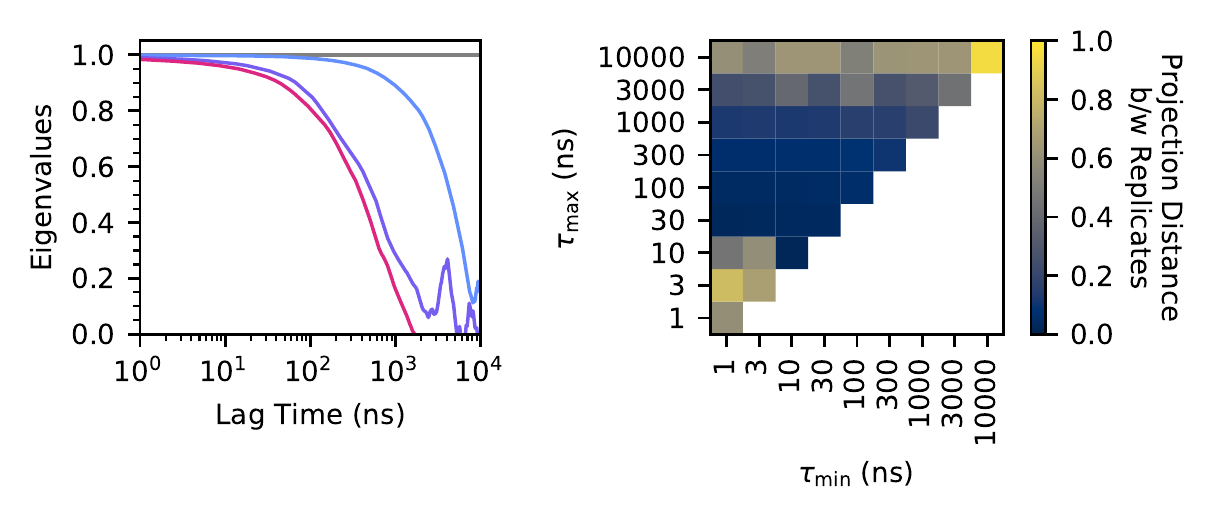}
\caption{Nonlinear IVAC results for the $125\ \mu \text{s}$ villin headpiece trajectory. 
(left) Estimated eigenvalues of the transition operator.
\added{(right) Root mean square projection distance between 10 replicates of nonlinear IVAC at the specified values of $\taumin$ and $\taumax$.}
}
\label{fig:villin_variability}
\end{figure*}

% The villin data set used in our analysis is the largest available that is directly suitable for use in IVAC. 
% Therefore, comparing IVAC results to the eigenfunctions of the transition operator is impossible for this problem.
% Here, we describe the protocols we use for selecting parameters,
% which help to ensure that IVAC is robust to estimation error.
% that are motivated by theory
% and check the consistency of results across multiple realizations of the neural network training procedure.

Here, we describe the protocols we use for selecting IVAC parameters.
\added{
By establishing clear protocols,
we help ensure that IVAC performs to the best of its ability,
providing robust eigenfunction estimates
even in a high-dimensional setting with limited data.
}

Our first protocol is to evaluate the condition number
for the subspace of eigenfunctions
that we are estimating.
This protocol is motivated by the theoretical error analysis in \citet{vacpaper1}, where
we showed that spectral estimates are less sensitive
to estimation error for a well-conditioned subspace.
To ensure that we are estimating a well-conditioned subspace, 
we first use IVAC to estimate eigenvalues for the transition operator.
We then identify a subspace of eigenfunctions $\eta_1, \eta_2, \ldots, \eta_k$ that is separated 
from all other eigenfunctions by a large spectral gap
$\hat{\lambda}_k^\tau - \hat{\lambda}_{k+1}^\tau$.

For the villin data, we choose the subspace consisting only of the constant eigenfunction $\eta_1 = 1$
and the first nontrivial eigenfunction $\eta_2$.
This is a well-conditioned subspace
with a minimum condition number
\begin{equation}
\min_\tau \kappa^\tau = \min_{\tau} \left(\hat{\lambda}_2^\tau - \hat{\lambda}_3^\tau\right)^{-1} = 1.6.
\end{equation}

\begin{figure*}
\centering
\includegraphics[scale = .8]{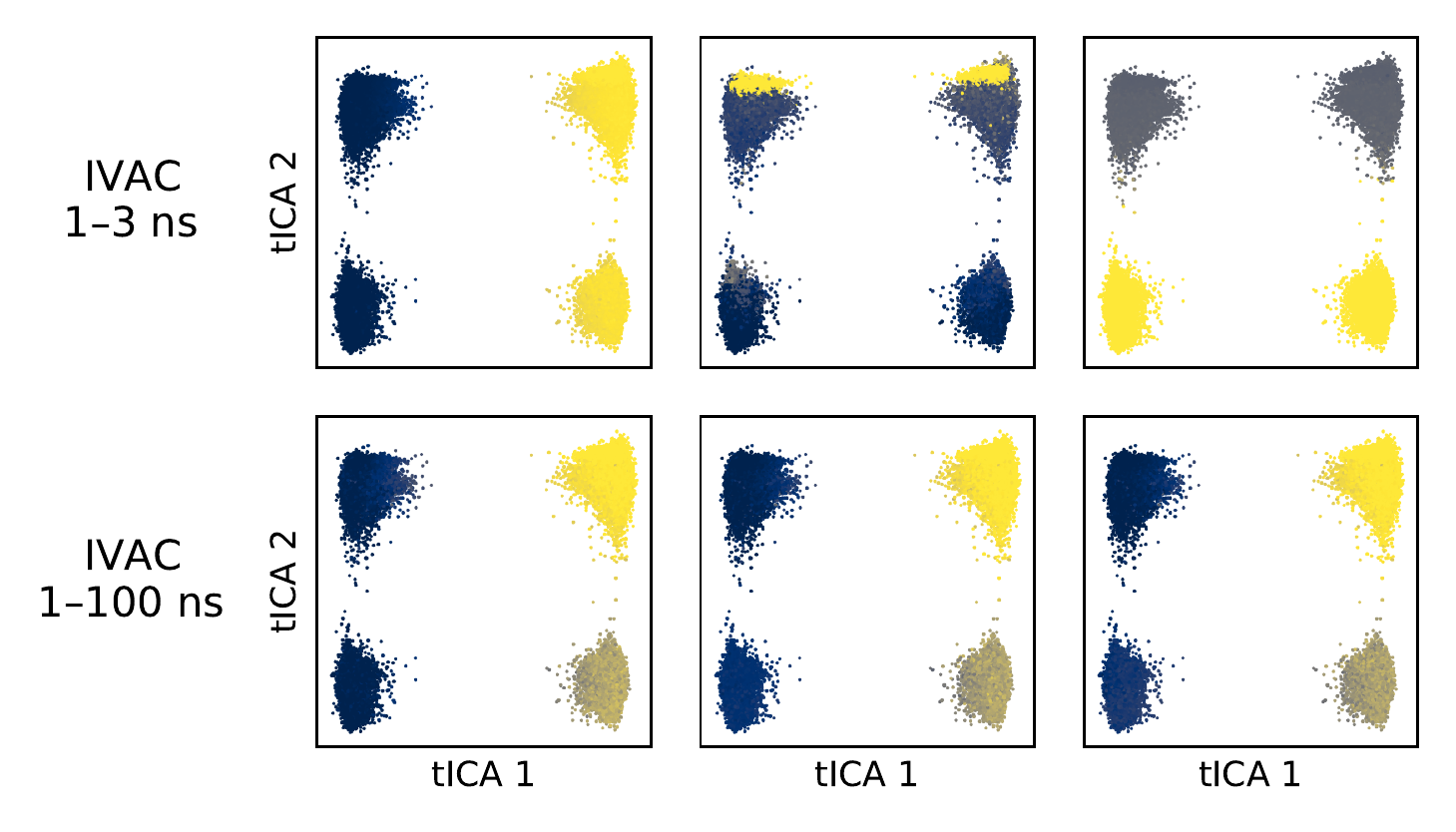}
\caption{Nonlinear IVAC results plotted on the first two time-lagged independent component analysis (tICA) coordinates.
(top) IVAC with a $1-3\ \text{ns}$ integration window and three different random seeds.
(bottom) IVAC with a $1-100\ \text{ns}$ integration window and three different random seeds.
\label{fig:villin_evecs}
}
\end{figure*}

Our second protocol
for ensuring robustness is to check that eigenfunction estimates
remain consistent
% across replications
% when the neural network is randomly initialized
% and optimized with randomly drawn subsets.
\added{
when the random seeds used in  initialization and subsampling are changed.
We train ten nonlinear IVAC neural networks
and quantify the inconsistency in the results
using the root mean square projection distance
between eigenspace estimates from different runs.
}
% \begin{equation}
% \sqrt{\frac{1}{N (N - 1) / 2} \sum_{i=1}^N \sum_{j=i+1}^N d(\mathcal{U}_i, \mathcal{U}_j)^2}
% \end{equation}
The results of this calculation are plotted in Figure \ref{fig:villin_variability}
across a range of $\taumin$
and $\taumax$ values.
The results for VAC
appear along the diagonal of the plot in Figure \ref{fig:villin_variability}, 
corresponding to the case $\taumin = \taumax$.

\added{
Figure \ref{fig:villin_variability} reveals problems with consistency for both IVAC and VAC, .
IVAC is robust to the choice of $\taumin$.
However,
% For IVAC, 
setting $\taumax < 30\ \text{ns}$ 
}
or $\taumax > 300\ \text{ns}$ 
leads to poor consistency.
If we train the neural network
with these problematic $\taumax$ values,
then solutions can look very different depending on the random seeds that are used for optimizing.
% Internal consistency is not just a problem with IVAC, however.
% We see the internal consistency is just as problematic for VAC as for IVAC.
With VAC, setting $\tau < 10\ \text{ns}$ or $\tau > 300\ \text{ns}$
would lead to inconsistent results.

% In the analysis of the alanine dipeptide data,
% IVAC performs well as long as the window of lag times includes the optimal VAC lag times.
% In the analysis of the villin data, however,
% the error levels are higher so
% we have to be more careful when choosing $\taumax$.
% If $\taumax$ is too high or too low, we find that the internal consistency suffers.
IVAC provides more flexibility to address the consistency issues compared to VAC,
since we can integrate over a range of lag times.
For the villin data, we choose to set $\taumin = 1\ \text{ns}$ and $\taumax = 100\ \text{ns}$.
For these parameter values, the consistency score is very good.
The typical projection distance between subspaces with different random seeds is just $0.05$. 
% representing $\arcsin\left(0.05\right)$ radians
% of separation in the estimates of the eigenfunction $\eta_2$
Moreover, $1-100\ \text{ns}$ is a wide range of lag times, helping to ensure that optimal or near-optimal VAC lag times are included in the integration window.

To help explain why the consistency 
%of IVAC 
is so poor for small $\taumax$ values,
we present in Figure \ref{fig:villin_evecs} a set of IVAC solutions obtained with an integration window of $1-3\ \text{ns}$ and three different random seeds.
We see that all three solutions identify clusters in the data,
but the clusters are completely different in the three cases.
We conjecture that IVAC is randomly fitting three different eigenspaces.
This is supported by the eigenvalue plot in Figure \ref{fig:villin_variability},
which shows that three nontrivial eigenvalues of the transition operator lie close together over the $1-3\ \text{ns}$ time window, 
making it possible that eigenspaces are randomly misidentified by IVAC.

In contrast to the inconsistent results obtained with an integration window of $1-3\ \text{ns}$,
we obtain more reasonable results
with an integration window of $1-100\ \text{ns}$.
As shown in Figure \ref{fig:villin_evecs},
the IVAC solutions are nearly identical regardless of the random seed.

\added{
In summary, we have proposed a robust procedure for approximating eigenfunctions of the villin headpiece system. 
We have chosen to approximate a well-conditioned eigenspace that is separated from other eigenspaces by a wide spectral gap. 
Moreover, we have ensured that IVAC results are consistent regardless of the random initialization and random drawn subsets used to train the neural net. 
Because of these protocols, the neural net estimates shown in Figure~\ref{fig:villin_evecs} reliably identify clusters in the trajectory data indicative of folded/unfolded states. 
}
% Moreover, the implied timescale estimate of $7.0\ \mu \text{s}$ versus is longer than the implied timescale estimate of $6.1\ \mu \text{s}$ that would be obtained using linear IVAC, indicating that a more accurate eigenfunction estimate can be obtained through the neural net ansatz.

\subsubsection{Robustness to overfitting}

In this section, we present results suggesting that nonlinear IVAC
is more robust to overfitting than nonlinear VAC.
This is crucial if the data set is too small for cross-validation.
% The added robustness can be crucial if there are insufficient data to construct a statistically meaningful validation data set, heightening the tendency to overfit. 
% and therefore the tendency toward overfitting is heightened.

\begin{figure}
\centering
\includegraphics{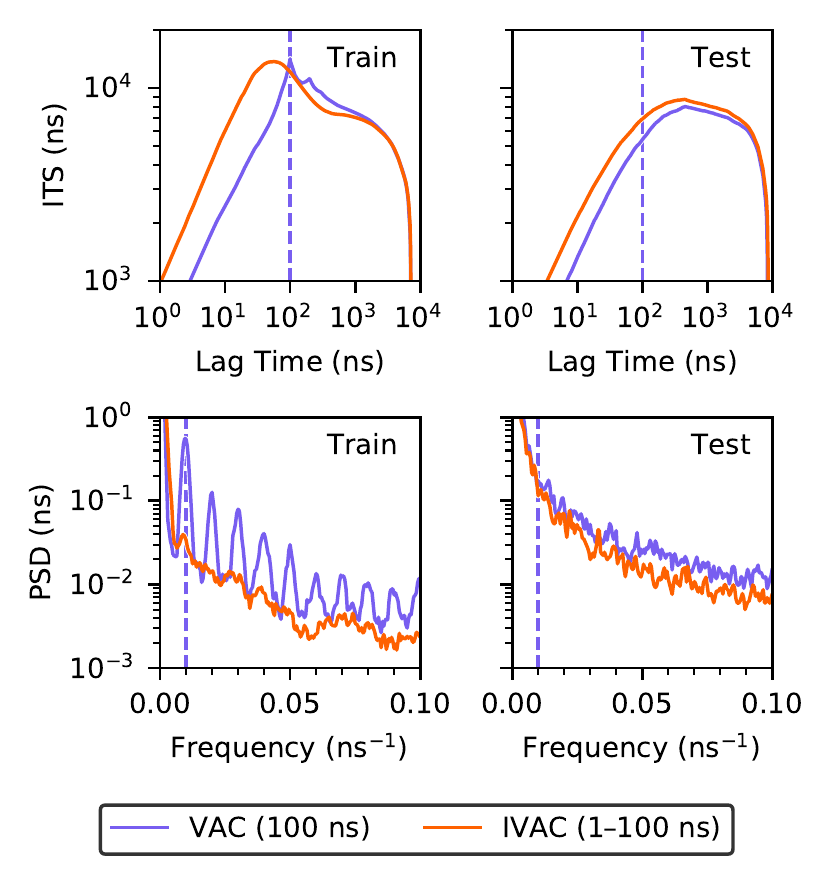}
\caption{Implied time scales (ITS) and power spectral densities (PSD) obtained with nonlinear IVAC and VAC with neural network basis functions applied to the villin headpiece data set. The VAC training lag time is marked by the dotted line in each panel.}
\label{fig:villin_artifacts}
\end{figure}

To identify the overfitting issue with small data sets,
we eliminate the early stopping and we train IVAC and VAC until the training loss stabilizes.
We calculate implied timescales by performing linear VAC on the outputs of the networks trained using IVAC and VAC, which we present in Figure \ref{fig:villin_artifacts}.

% As a first observation, we find that without early stopping
% both IVAC and VAC give lower-quality solutions.
% The implied timescale estimates for the validation data are lower than the implied timescale estimates for the training data.
% The discrepancy indicates that IVAC is fitting solutions to the noise in the data,
% not the underlying signals from the transition operator,
% and overfitting has occurred.
% The overfitting reinforces the utility of using a train-test split and applying early stopping
% whenever feasible.

We first compare the estimated implied timescales between the training and validation data sets.  For both algorithms, the implied timescales calculated on the training data are larger than those calculated on the validation data.  
This is clear evidence of overfitting.
However, we see that IVAC gives larger implied timescales on the validation data
compared to VAC.
In combination with the variational principle associated with the implied timescales, this suggests that IVAC is giving an improved estimate for the slow eigenfunctions.

Examining the implied timescales estimated on training data show further signs of overfitting.
The VAC implied timescale estimates for the training data exhibit sharp peaks at the training lag time that are absent in the implied timescale estimates of the validation data.
This suggests a hypothesis for the mechanism of overfitting:
with a sufficiently flexible approximation space, VAC is able to find spurious correlations between features that happen to be separated by $\tau$.
This explains the smaller peaks at integral multiples of the lag time, as features artificially correlated at $\tau$ will be correlated at $2 \tau$ as well.

To confirm our hypothesis, we 
%
% As seen in the figure, the impact of overfitting is worse for VAC
% than it is for IVAC.
% Notice that the VAC implied timescale estimates exhibit sharp peaks at integer multiples of the training lag time.
% These peaks appear only for the training data set, not the test data set,
% which suggests that the periodicity is an artifact of overfitting.
% To highlight this periodicity,
we plot the power spectral density (PSD) \cite{welch1967}
of the time trace of eigenfunction estimates
in Figure \ref{fig:villin_artifacts}.
The PSD confirms the existence of a periodic component in VAC results with a frequency at the inverse training lag time.
In contrast, IVAC does not exhibit such a periodic component.
In Figure \ref{fig:villin_artifacts}, we see that the $1-100\ \text{ns}$ integration window leads to implied timescale estimates
that depend smoothly on the data both for the training and the test data set.
The PSD shows no periodic components in the spectra for IVAC, providing further evidence that 
IVAC is comparatively robust
while VAC results can be very sensitive to the particular lag time that is used.

%% file: conclusion.tex
In this paper we have presented integrated VAC (IVAC), a new extension to the popular  variational approach to conformational dynamics (VAC).
By integrating correlation functions over a window of lag times,
IVAC provides robust estimates of the eigenfunctions of a system's transition operator.

To test the efficacy of the new approach, we compared IVAC and VAC results on two molecular systems.
First, we applied the spectral estimation methods to simulation data from the alanine dipeptide.
This is a relatively simple system that permits generation of extensive reference data for validating our calculations.
As we varied the lag time parameters and the amount of data available,
we observed the improved robustness of IVAC compared to VAC.
IVAC gives low-error eigenfunction estimates
even when the lag times range over multiple orders of magnitude.
In contrast, VAC requires more precise lag-time tuning to give reasonable results
% , especially in the low-data regime.

Next we applied IVAC to analyze a folding/unfolding trajectory for the villin headpiece.
These data contain relatively few 
%events of interest 
folding/unfolding events
despite pushing the limits of present computing technology.
For this application, we used a flexible neural network representation built on top of a rich feature set.
We presented a procedure for selecting parameters in IVAC
that helps lead to robust performance
in the face of uncertainty.
% Additionally,
% in this case, 
For the application to villin data,
we found that
VAC exhibited pronounced artifacts from overfitting when precautions were not taken to specifically prevent it, while IVAC did not.

Our work highlights the sensitivity of VAC calculations to error from insufficient sampling.
Examining our results on the villin headpiece, we see that regularization (here, by early stopping) and validation are crucial when running VAC with neural networks or other flexible approximation spaces.
With insufficient regularization or poor validation these schemes easily overfit.
Even for the alanine dipeptide example, where we employ a simple basis on a statistically well-conditioned problem, we see that VAC has a high probability of giving spurious results with insufficient data.

Integrated VAC addresses this problem by considering information across multiple time lags.
Future extensions of the work could further leverage this information.
For instance, employing a well-chosen weighting function within the integral in~\eqref{eq:Rayleigh_w_T_op} could further decrease hyperparameter sensitivity.
Additionally, future numerical experiments could point to improved procedures for selecting $\taumin$ and $\taumax$ values.
Finally, we could integrate over multiple lag times in other formalisms using the transition operator, such as schemes that estimate committors and mean-first-passage times \cite{thiede2019galerkin}.
% Finally, just as VAC generalizes MSM estimates of eigenfunctions, a related formalism generalizes MSM estimates of quantities central to rate theory, such as committors and mean-first-passage times \cite{thiede2019galerkin}.
% These calculations are more sensitive to choice of lag time than VAC, and it is possible that an integral formulation could resolve these issues.
These extensions would further strengthen the basic message of our work: combining information from multiple lag times
leads to improved estimates of the transition operator and its properties.

%% file: acknowledgments.tex
EHT was supported by DARPA grant HR00111890038.   
RJW was supported by the National Science Foundation through award DMS-1646339. CL, ARD, and JW were supported by the National Institutes of Health award R35 GM136381. JW was supported by the Advanced Scientific Computing Research Program within the DOE Office of Science through award DE-SC0020427.
The villin headpiece data set was provided by D.E. Shaw Research.
Computing resources where provided by the University of Chicago Research Computing Center.